\documentclass[12pt]{article}
\usepackage{amsbsy}
\usepackage{amsfonts}
\usepackage{amsmath}
\usepackage{amssymb}
\usepackage{apacite}
\usepackage{setspace}
\usepackage{graphicx}
\usepackage{bm,multicol}
\usepackage{xfrac}
\usepackage[english]{babel}
\usepackage{natbib}
\usepackage[T1]{fontenc}
\usepackage[utf8]{inputenc}
\usepackage{authblk}
\usepackage{xcolor}

\newcommand{\newc}{\newcommand}
\newc{\N}{\mbox{N}}
\newc{\1}{\bf{1}}

\usepackage{graphicx}

\usepackage{multirow}

\begin{document}
\title{Bayes Factor Hypothesis Testing in Meta-Analyses: Practical Advantages and Methodological Considerations
}

\author{Joris Mulder$^{*}$ \& Robbie C.M. van Aert$^{\dagger}$}
\date{%
    Dept. of Methodology and Statistics, Tilburg University\\
    \bigskip
    $^*$\small{j.mulder3@tilburguniversity.edu}\\
    $^{\dagger}$r.c.m.vanaert@tilburguniversity.eduå
}

\maketitle

\begin{abstract}
Bayesian hypothesis testing via Bayes factors offers a principled alternative to classical p-value methods in meta-analysis, particularly suited to its cumulative and sequential nature. Unlike commonly reported p-values for standard null hypothesis significance testing, Bayes factors allow for quantifying support both for and against the existence of an effect, facilitate ongoing evidence monitoring, and maintain coherent long-run behavior as additional studies are incorporated. Recent theoretical developments further show how Bayes factors can flexibly control Type I error rates through connections to e-value theory. Despite these advantages, their use remains limited in the meta-analytic literature. This paper provides a critical overview of their theoretical properties, methodological considerations—such as prior sensitivity—and practical advantages for evidence synthesis. Two illustrative applications are provided: one on statistical learning in individuals with language impairments, and another on seroma incidence following post-operative exercise in breast cancer patients. New tools supporting these methods are available in the open-source R package BFpack.

\end{abstract}


\section*{Highlights}
\subsection*{What is already known}
\begin{itemize}
\item Meta-analysis methods are widely used to combine effect sizes across studies, typically within a traditional frequentist framework.
\item These methods face challenges in hypothesis testing because the cumulative nature of meta-analyses inherently induces multiple testing issues.
\item Bayes factors provide an alternative that directly quantify the evidence between hypotheses, and allow for natural evidence accumulation.
\end{itemize}
\subsection*{What is new}
\begin{itemize}
\item The paper compares Bayes factor testing with classical significance testing in meta-analyses, clarifying their conceptual and methodological differences.
\item It presents five Bayes factor models for evidence synthesis, illustrated using the standard single-effect-size meta-analysis setup.
\item The paper discusses prior specification, including priors for the (nuisance) between-study heterogeneity.
\item It highlights the link between Bayes factors and e-values as a means for flexible classical error control in cumulative meta-analyses.
\item All methods are implemented in the R package \texttt{BFpack}.
\end{itemize}
\subsection*{Potential impact}
\begin{itemize}
\item The overview aims to guide researchers in selecting suitable evidence synthesis methods and promote flexible, statistically robust Bayesian approaches for hypothesis testing in (cumulative) meta-analyses.
\end{itemize}

\section{Introduction}
Meta-analysis refers to the statistical methodology used for combining independent studies addressing the same research question. the approach improves the precision of results, combines the available evidence, and it may also resolve controversies when contradicting conclusions are drawn in multiple studies \citep{deeks2023}. Given the available published studies, a meta-analyst is often interested in estimating the magnitude of a global effect and its statistical uncertainty. Instead of or next to estimation, the focus of a meta-analyst can be on testing whether the effect is equal to a specific value, typically zero \citep[e.g.,][]{higgins2009,higgins2011sequential,rice2018re}. This is, for instance, of interest if the goal is to answer whether a treatment is beneficial on average. 

Over the last decades, Bayesian estimation methods have become increasingly popular \citep[e.g.,][]{smith1995bayesian,higgins2009,turner2015,hackenberger2020, schmid2021}. These methods may be more accurate in case of few studies by not relying on large sample theory and the possibility to include external information in the prior distribution \citep{friede2017a, rhodes2016}. If the information in the data dominates the prior, classical and Bayesian estimations methods behave similarly. When the goal is to test the effect on a specific value (or specific range), classical significance-based testing is most common using classical two-sided p-values. The test may also be executed by evaluating whether the specific (null) value falls inside the classical confidence interval (CI) or inside the Bayesian credible interval (CrI). Another way to test hypotheses in meta-analyses is using the Bayes factor, a Bayesian criterion for hypothesis testing \citep{Jeffreys,Kass:1995}. Bayes factors possess fundamentally different properties from significance-based tests. This Bayesian criterion may be particularly useful for meta-analyses where statistical evidence accumulates across multiple studies. While meta-analyses are typically modeled as combining independent studies, in practice earlier findings often influence whether subsequent studies are conducted, meaning that true independence rarely holds \citep[e.g.,][]{ter2019accumulation}. This implicit sequential dependence further motivates the use of Bayes factors, which remain valid under such cumulative accumulation of evidence. As the meta-analysis community is less familiar with this alternative methodology, this paper aims to provide a critical overview of theoretical properties, methodological considerations and practical advantages of this methodology. Table \ref{tab_diffs} summarizes key conceptual and practical differences between classical p-value testing and Bayes factor testing, which we elaborate upon below.

\begin{table}[thp]
\caption{Summary of differences between classical p-value and Bayes factor testing.}
\begin{center}
\hspace*{-2cm}
\begin{tabular}{lcc}
  \hline
 & Classical p-value testing & Bayes factor testing \\
\hline
0. Goal & Make a dichotomous & Quantify relative evidence\\
& decision while controlling  & given the available infor-\\
& the type I error rate  & mation\\[3mm]
1. Necessary to  & Significance level $\alpha$ & Meaningful prior for\\
initialize the test & & the key parameter.\\[3mm]
2. Computation & Usually via large sample theory & Usually via numerical algorithms\\[3mm]
3. Provide evidence in & Via additional equivalence & Natural property. No\\
favor of $\mathcal{H}_0$ &  testing \citep{lakens2017equivalence}  & additional steps required \\
&  & \citep[e.g.][]{Rouder:2009}\\[3mm]
4. Distinguish between & Via additional power & Natural property.  No\\
absence of evidence and & studies under $\mathcal{H}_1$ & additional steps required \\
evidence of absence & \citep{hoenig2001abuse} & \citep[e.g.][]{dienes2014using}  \\[3mm]
5. Monitor test in& In a restrictive, planned & Natural property depending\\
on-going meta-analyses &   manner to control type I & on the prior \citep{de2021optional} \\
                       &  error rate \citep{higgins2011sequential} &  \\ [3mm]
6. Consistent testing & Always when $\mathcal{H}_1$ is true; & Natural property when \\
behavior in the long run & inconsistent when $\mathcal{H}_0$ is true & either $\mathcal{H}_0$ or $\mathcal{H}_1$ is true \\[3mm]
7. Able to test order-  & Limited to nested hypotheses & For nested and nonnested \\
constrained hypotheses  & \citep{Silvapulle:2004} &  hypotheses \citep[e.g.][]{Hoijtink:2011} \\
\hline
\end{tabular}
\end{center}
\label{tab_diffs}
\end{table}


The goal of classical p-value testing is to make a dichotomous decision while controlling the type I error rate at a particular prespecified $\alpha$-level. This is in contrast with the goal of the Bayes factor which quantifies the relative evidence in the available published studies between the hypotheses via the ratio of the so-called marginal likelihoods of the available data under the respective hypotheses, i.e., $\mathcal{H}_1$ and $\mathcal{H}_0$. Mathematically, the Bayes factor is defined by:
\begin{eqnarray}
\label{BF10}
B_{10}(y_{1:k}) &=& \frac{p(y_{1:k}|\mathcal{H}_1)}{p(y_{1:k}|\mathcal{H}_0)}, 
\end{eqnarray}
where $y_{1:k}$ denotes the available effect sizes from studies 1 to $k$. Given the interpretation as a measure of relative evidence, and because a meta-analysis aims to synthesize evidence across studies, the use of Bayes factors for meta-analyses is sometimes termed Bayesian evidence synthesis \citep[e.g.,][]{scheibehenne2016,ly2019replication,klugkist2023bayesian}.

As can be seen from \eqref{BF10}, the Bayes factor is a type of likelihood ratio. Unlike the classical likelihood ratio test statistic, which is computed at the maximum likelihood estimates of the unknown parameters \citep[][Ch. 8]{casella2024statistical}, the marginal likelihoods are computed as weighted averages of the likelihood weighted according to the prior distributions of the unknown parameters under the hypotheses \citep{Jeffreys,Kass:1995}. Therefore, the Bayes factor is sensitive to the choice of the prior: its outcome is only meaningful when the chosen priors are meaningful, especially for the parameter that is tested. In Bayesian estimation of meta-analysis models on the other hand, the prior plays a considerably smaller role if `vague', weakly informative, or noninformative priors are used\footnote{The prior for the between-study heterogeneity parameter can be important in Bayesian estimation under random effects meta-analysis models especially for a small number of studies \citep{rhodes2015, turner2015}}. When testing hypotheses using the Bayes factors, extremely vague priors should not be used for the effect that is tested. Such priors cover unrealistically large effect sizes, often resulting in Bayes factors that are unrealistic quantifications of the relative evidence between the hypotheses.

While this prior sensitivity is sometimes cited as a limitation \citep[e.g.,][]{higgins2009}, classical testing approaches also require subjective inputs, such as defining a minimal effect of interest for equivalence testing or choosing a plausible effect size for power analysis \citep{lakens2017equivalence, hoenig2001abuse}. Hence, both statistical approaches demand thoughtful specification of the alternative hypothesis \citep{rouder2016there}. Because of the importance of prior specification, this paper elaborately discusses this topic under various meta-analysis models.

\begin{figure}[t]
\begin{center}
\includegraphics[height=2.5cm]{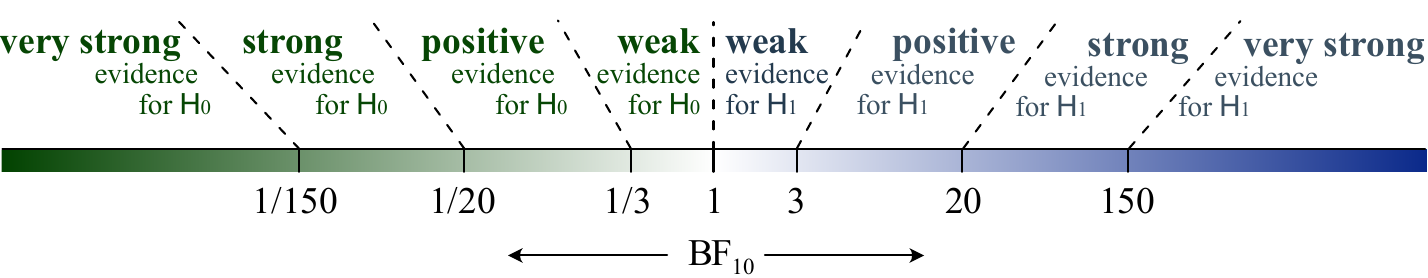}
\caption{Interpreting the evidence on a continuous (log) scale. The qualitative categories can be found in \cite{Kass:1995}. Visualization of the colored bar from \cite{mulder2024bayesian}.}
\label{fig_BFscale}
\end{center}
\end{figure}

Once priors are specified, computing the Bayes factor usually requires intensive numerical methods, unlike classical p-value tests that rely on simpler large-sample calculations. To interpret a Bayes factor, Figure \ref{fig_BFscale} displays the relative evidence between the hypotheses on a continuous scale. For example, a Bayes factor of $B_{10}=15$ implies that the data were 15 times more plausible under the alternative $\mathcal{H}_1$ than under the null $\mathcal{H}_0$, implying clear evidence in favor of $\mathcal{H}_1$. On the other hand, a Bayes factor of, say, $B_{10}=0.14$, implies that we obtained positive evidence in favor of $\mathcal{H}_0$ because $B_{01}=1/B_{10}\approx 7.1$ implying that the data were 7.1 times more plausible under $\mathcal{H}_0$. This illustrates that Bayes factors allow evidence quantification in favor of a null hypothesis. Depending on the field of research, this natural property may be particularly important because null hypotheses may often be true \citep{johnson2017reproducibility}. 


Moreover, if the Bayes factor is close to 1, this would imply absence of evidence towards any of the two hypotheses (Table 1; Dienes, 2014). This illustrates that Bayes factors have the natural ability to distinguish between absence of evidence (i.e., an underpowered analysis when $B_{10}\approx 1$) and evidence of absence (i.e., evidence in favor of the null when $B_{01}\gg1$). To assess whether the test was underpowered using classical testing, additional power analyses would have been required. However, when power analyses have not been executed before the analysis, post-experimental power analyses are not without problems \citep{hoenig2001abuse}.

Figure~\ref{fig_BFscale} shows qualitative bounds for interpreting Bayes factors, as proposed in the literature \citep[e.g.,][]{Jeffreys,Kass:1995}, which serve primarily as a guide for researchers less familiar with the concept and should not be applied rigidly. While Bayes factors naturally provide a graded measure of evidence, they can also be compared against a threshold to make dichotomous decisions, potentially controlling classical type I error rates even in on-going meta-analyses where studies are evaluated sequentially, regardless of the stopping rules or data-collection decisions applied in previous studies \citep{safetesting,de2021optional,ter2019accumulation,rouder2014optional}. Classical tests, by contrast, require careful pre-planning for such designs \citep[see Table~\ref{tab_diffs};][]{higgins2011sequential}. This type I error control for Bayes factors in sequential settings is enabled by recent advances in ``e-value theory'', which supports `safe anytime-valid inference'—a relatively novel statistical framework ensuring that statistical conclusions remain valid regardless of when data collection or analysis stops, without requiring careful pre-planning \citep{ramdas2024hypothesis, ly2024tutorial, safetesting}. Moreover, Bayes factors are statistically consistent, with evidence accumulation towards the true hypothesis as the number of studies grows, whereas classical tests remain inconsistent due to a persistent chance of rejecting a true null at the pre-specified $\alpha$-level.

Finally, Bayes factors are relatively flexible for testing more complex hypotheses involving combinations of equality and order constraints on multiple parameters. Such hypotheses can reflect more precise scientific expectations regarding the specific relationships between the parameters (such as order constraints between group means). Though not very common, they have been used for meta-analytic applications \citep[e.g.][]{kuiper2013combining,van2024bayesian}. Although p-values are also available for testing such hypotheses, the class of order-constrained hypotheses that can be tested is limited \citep[e.g., only nested hypotheses can be tested against each other;][]{Silvapulle:2004}.

To guide researchers interested in using Bayes factors for hypothesis testing in meta-analyses, we begin with a published example that motivates our work (Section~\ref{SectionExample}). Section~\ref{SectionMetaModels} introduces five meta-analytic models, focusing on the standard framework of normally distributed effect sizes with known error variances and independent contributions per study. This standard setup was chosen for accessibility and because normal models are most often used. Naturally, it is generally advisable to use exact models when appropriate (e.g., a logistic model for binary data). Throughout the paper, we focus on the (most common) two-sided hypothesis test. Section 4 discusses prior specification for the (average) effect size, which may reflect the standardized mean difference, log odds ratio, or Fisher-transformed correlation, with brief remarks on priors for between-study heterogeneity. Section~5 outlines how to compute Bayes factors for the five models. Section~6 connects Bayes factors to e-values, highlighting their suitability for sequential meta-analysis. Section~7 provides a synthetic illustration, and Section~8 applies Bayes factors to two real meta-analyses. We conclude with a discussion, and note that the R package \texttt{BFpack} \citep{mulder2021bfpack} has been extended to support several of the Bayes factor tests presented here.

\section{Motivating illustration}\label{SectionExample}

\begin{figure}[thp]
\begin{center}
\includegraphics[width=8cm]{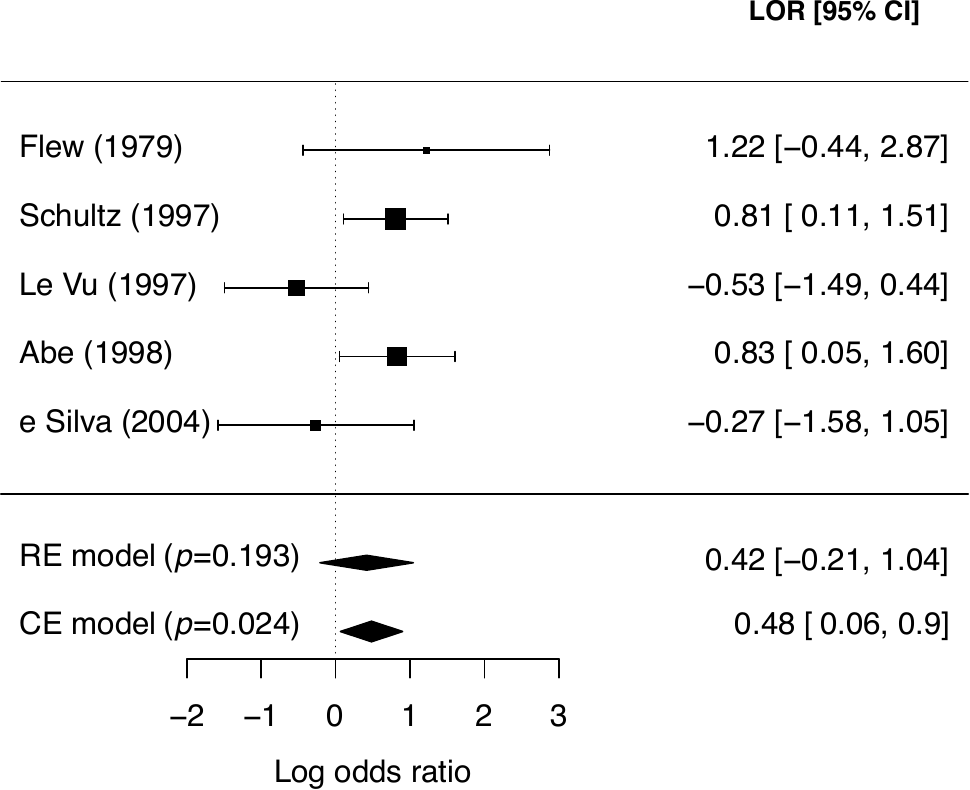}
\caption{Forest plot for the meta-analysis of McNeely et al. (2010). LOR is the log odds ratio, and RE and CE refer to random effects and common effect, respectively.}
\label{fig_mcneely0}
\end{center}
\end{figure}

\cite{mcneely2010} presented a meta-analysis on the incidence of seroma when patients start exercising within or after three days following a breast cancer surgery. Five studies are included in this meta-analysis where patients were assigned to an early or delayed exercise condition in each study. The outcome variable was the occurrence of seroma. Thus, a log odds ratio was the effect size measure of interest. A log odds ratio larger (smaller) than zero indicates that seroma is more (less) likely to appear in this early period compared to delayed exercise condition. The data (including the publication years of the studies in chronological order) and the corresponding 95\%-CIs for this meta-analysis are presented in the forest plot in Figure \ref{fig_mcneely0}.

We can formulate the following two-sided test for this application:
\begin{eqnarray*}
\mathcal{H}_0 &:& \text{``The average effect is zero''}\\
\mathcal{H}_1 &:& \text{``The average effect is nonzero (either positive or negative)''.}
\end{eqnarray*}
If the conventional significance level of 5\% were applied, the null hypothesis would have been rejected under the common effect (CE) model but not under the random effects (RE) model. The null hypothesis of no between-study heterogeneity was not rejected based on the Q-test ($Q(4) = 7.765$, $p = .101$). However, this test is known to have low power when only a few studies are available \citep{hardy1998}. Given the considerable heterogeneity among the effects observed in the individual studies, a more conservative approach—such as the RE model used by \cite{mcneely2010}—is likely more appropriate. Under this model, there is insufficient evidence to reject the null hypothesis regarding the average effect. However, the outcome of this classical test does not clarify whether the nonsignificant result reflects a true null effect or simply a lack of statistical power—especially since \cite{mcneely2010} did not report a power analysis.

Due to this statistical uncertainty, a research group may be motivated to conduct a new study on the incidence of seroma. Once published, however, updating the meta-analysis using classical significance testing introduces a challenge: what significance level ($\alpha$) should be used when testing the null hypothesis a second time? Re-testing naturally inflates the overall significance level. Ignoring this multiple testing problem undermines the core assumption of a fixed sample size that underlies classical p-value-based inference.

Bayes factors offer a compelling alternative in such settings. First, they can distinguish between a lack of evidence (i.e., an underpowered study, leading to a Bayes factor close to 1) and evidence favoring the null hypothesis (i.e., a Bayes factor indicating substantial support for the null; Figure \ref{fig_BFscale}). Second, Bayes factors allow for straightforward updating of the relative evidence for competing hypotheses as new studies become available. Furthermore, recent advances in e-value theory \citep{ramdas2024hypothesis,grunwald2024safe,hendriksen2021optional,de2021optional} make it possible to maintain Type I error control without relying on subjective priors, thus preserving desirable frequentist properties while still benefiting from Bayesian updating.

Finally note that a meta-analyst may consider reporting the Bayesian probability that the average effect is positive, particularly if this direction is expected. Such probabilities behave similarly to one-sided p-values however \citep{marsman2017three}, and thus face similar challenges as two-sided p-values in the context of hypothesis testing.

\section{Statistical models for Bayesian evidence synthesis}\label{SectionMetaModels}
Depending on the application, different meta-analysis models can be used. The current section gives a brief overview of three traditional and two more recent hybrid meta-analysis models. Bayes factor tests will be discussed under these models in subsequent sections. 

\subsection{Traditional meta-analysis models} 
\subsubsection{Common effect model}
Under the common effect (CE) model, the key parameter, denoted by $\theta$, is assumed to be common under all studies. In order for this assumption to hold, the conditions under which the data are collected under every study need to be (practically) identical, such as replication studies in psychology \citep{schmidt2009shall} or series of randomized clinical trials by the same researchers, for patients from the same population, and testing exactly the same treatment \citep{borenstein2010}.

For study $i$, for $i=1,\ldots,k$, we denote the estimated effect size by $y_i$, its (assumed known) standard error by $\sigma_i$, and the sample size in the study by $n_i$. A Gaussian (normal) error is assumed for the study-specific estimate and the corresponding standard error resulting in the following synthesis model:
\begin{equation}
\label{commoneffect}
\mathcal{M}_{CE}: y_i \sim N(\theta,\sigma^2_i), \text{ for study $i=1,\ldots,k$.}
\end{equation}
Under the CE model, we consider the most common test in statistical practice of a precise null hypothesis, which assumes that the mean effect is zero, against a two-sided null hypothesis test, which assumes that the mean effect is nonzero, i.e.,
\begin{eqnarray*}
\mathcal{H}_0&:&\theta = 0\\
\mathcal{H}_1&:&\theta \not = 0.
\end{eqnarray*}

\subsubsection{Random effects model}
Under the random effects (RE) model, the effects $\theta_i$ are assumed to be heterogeneous across studies. This heterogeneity may be caused by (slightly) different conditions under which the data were collected across studies or (slightly) different populations that were considered under the different studies. The RE model is generally the preferred model since researchers deem the assumption of a common effect unrealistically restrictive in most applications. Moreover, the RE model is also often preferred, because it allows for drawing inference for the distribution of true effects whereas the CE model is restricted to drawing inference to only the included studies in a meta-analysis.

A normal distribution is assumed for the study-specific effects where the mean $\mu$ quantifies the average (global) effect across studies and the standard deviation $\tau$ quantifies the between-study heterogeneity in true effect size. Similar as the CE model, normally distributed errors are assumed for the study specific estimates. The RE model can then be formulated as
\begin{eqnarray}
\label{randomeffects} \mathcal{M}_{RE}:\left\{\begin{array}{ccl}
y_i &\sim & N(\theta_i,\sigma_i^2)\\
\theta_i &\sim & N(\mu,\tau^2).
\end{array}
\right.
\end{eqnarray}
In this model, the study-specific true effects (i.e., $\theta_i$) are often treated as nuisance parameters which can be integrated out. The marginalized RE model can then be equivalently written as
\begin{eqnarray}
\label{randomeffects2} \mathcal{M}_{RE}:
y_i &\sim & N(\mu,\sigma_i^2+\tau^2),\text{ with $\tau^2>0.$}
\end{eqnarray}
Under the RE model, we generally test whether the global effect $\mu$ is zero or not i.e.,
\begin{eqnarray}
\label{hyporanef}\mathcal{H}_0&:&\mu = 0\\
\nonumber \mathcal{H}_1&:&\mu \not = 0.
\end{eqnarray}

\subsubsection{Fixed effects models}\label{fixedeffectsmodel}
Similarly as the RE model, and unlike the CE model, the fixed effects (FE) model also assumes heterogeneous effects across studies. However rather than specifying a distribution for the between-study heterogeneity, whose parameters are estimated from the data, a fixed effects approach is considered without a multilevel structure. The parameters of interest are the true effect sizes of the studies. Although less common, the FE model has been used in meta-analytic applications \citep{rice2018re,kuiper2013combining,klugkist2023bayesian,van2024bayesian,vanLissa2024} and when aggregating evidence across respondents \citep{stephan2007comparing,regenwetter2018heterogeneity,klaassen2018all}\footnote{Although these references used different terminologies for this model, we use the term `fixed effects model', following \cite{rice2018re}, to clearly distinguish from the CE model.}. Moreover, the fixed effects model can also be seen as a common effect model that is extended to a meta-regression model by the inclusion of a moderator that has a unique value for each study. According to this meta-regression model, each study also has its own true effect size

Again Gaussian errors are assumed for the study-specific effect size estimates. The FE model can then be formulated as
\begin{equation}
\mathcal{M}_{FE} : y_i \sim N(\theta_i,\sigma_i^2).
\label{fixefmodel}
\end{equation}
Because of the absence of a parameter for a global effect, the null hypothesis assumes that all study-specific effects are zero or not \citep{rice2018re}, i.e.,
\begin{eqnarray*}
\mathcal{H}_0&:&\theta_1=\ldots=\theta_k=0,\\
\mathcal{H}_1&:&\text{not $\mathcal{H}_0$},
\end{eqnarray*}
where the alternative assumes that at least one constraint under $\mathcal{H}_0$ does not hold. Hence, this null hypothesis fundamentally differs from the null hypothesis under the CE and RE model where the average effect across studies is assumed to be zero while in the FE model, the null assumes that all study-specific effects are assumed to be zero. Consequently, the null is extended with every newly included study. 

It has been argued that a FE model has the advantage that it can be used when the study designs and/or measurement levels of the key variables (highly) vary across studies \citep{kuiper2013combining,klugkist2023bayesian}. The argument is that we are not combining effect sizes, which may then have (highly) different scales, but rather combining statistical evidence regarding the hypotheses when computing Bayes factors. However, the relative evidence between hypotheses as quantified by the Bayes factor is directly affected by the observed effect size and its uncertainty (via the likelihood). Therefore, the appropriateness of combining evidence from different studies with highly different designs, e.g., studies with reported effect sizes based on both dichotomous and continuous outcomes, or studies with an experimental design or observational design, should be carefully assessed by substantive experts. Moreover, prior specification may be more challenging when considering effect sizes having fundamentally different scales.


\subsection{Hybrid effects model}
To apply a traditional meta-analysis model, a dichotomous decision needs to be made whether to assume between-study homogeneity (i.e., the CE model) or between-study heterogeneity (i.e., the RE model, which is much more common than the FE model). This comes down to the following model selection problem:
\begin{eqnarray}
\label{modelselect}\mathcal{M}_{CE}&:&\tau^2 = 0\\
\nonumber \mathcal{M}_{RE}&:&\tau^2 > 0.
\end{eqnarray}
Although the \textit{Q}-test \citep{cochran1954} can be used for testing whether the data are homogeneous, it is not recommended for model selection since it may have low statistical power depending on the number of studies included in the meta-analysis, the sample size of the studies, and the true between-study heterogeneity \citep{viechtbauer2007confidence, borenstein2010}. This implies potentially large error rates when choosing either one of the two possible models.

Specifically, when an incorrect CE model is employed, the standard error of the key parameter will be underestimated. In a classical significance test, this would results in inflated type I error rates, and in a Bayesian evidence synthesis, this would result in an overestimation of the evidence for the true hypothesis. On the other hand, when an incorrect random effects model is employed, the standard error of the key parameter will be overestimated unless $\tau^2$ is estimated as zero. In a classical significance test, this results in an underpowered test, and in Bayesian evidence synthesis, this would result in an underestimation of the evidence for the true hypothesis. Thus, when there is considerable statistical uncertainty regarding the true underlying model and there are no theoretical reasons for favoring one model over the others, it is useful to employ a statistical model that encompasses both the CE and RE model to avoid a potentially error-prone dichotomous decision resulting in unreliable quantifications of the statistical evidence \citep{gronau2021primer,van2022bayesian}. We shall refer to the class of models that encompass both the CE model and the random effects model as hybrid effects models. To our knowledge, two hybrid effects models have been proposed in the literature which we discuss below. Appendix \ref{AppHybrid} discusses some statistical differences.

\subsubsection{Marginalized random-effects meta-analysis (marema) model}
The marginalized random-effects meta-analysis (marema) model \citep{van2022bayesian} is closely related to the RE model with the exception that it also allows for the possibility for excessive homogeneity implying less variability across studies than would be expected by chance. The marema model can be written as
\begin{equation}
\mathcal{M}_{marema}: y_i \sim N(\mu,\sigma_i^2+\tau^2),\text{ with $\tau^2 > -\sigma_{\min}^2$},
\end{equation}
where the lower bound of $\tau^2$ depends on the smallest sampling variance of the included studies, $\sigma_{\min}^2=\min_i \{\sigma^2_i\}$. Thus, under the marema model, $\tau^2$ can attain negative values. A negative $\tau^2$ implies that the between-study heterogeneity is smaller than expected by chance.

Although this property may seem unnatural at first sight, this setup has various advantages. First, the model allows a simple check of whether between-study heterogeneity is present via the posterior probability that $\tau^2>0$ holds (this simple Bayesian measure can also be used as an alternative to the $Q$-test as it does not rely on large sample theory). Second, the model allows noninformative improper priors for $\tau^2$ (both for estimation, e.g., whether $\tau^2>0$, and for testing the global mean $\mu$ using a Bayes factor). Thereby, the model simplifies the (challenging) choice of the prior for $\tau^2$. Third, the model enables researchers to check for extreme between-study homogeneity (less than expected by chance), which may indicate strong correlation between studies, extreme bias, or potential fraud \citep{ioannidis2006}. This can be checked via the posterior probability that $\tau^2<0$. Fourth, as mentioned earlier, the models avoid the need to make a dichotomous decision between the CE model and the RE model but instead naturally balances between these models depending on the between-study heterogeneity that is present. Finally note that negative variances are very common in latent variable models (such as the RE model). In the factor analysis literature, these are known as `Heywood cases', which often indicate model misspecification \citep{kolenikov2012testing}. In our current setup, this implies that the RE model would be inappropriate given the available data. Marginalized RE models have also been advocated for various other statistical problems \citep[e.g.,][]{mulder2019bayes,fox2017bayes,nielsen2021small}.

Under the marema model, the hypothesis test on the global effect will be the same as under the RE model, i.e., 
\begin{eqnarray*}
\mathcal{H}_0:\mu = 0,\\
\mathcal{H}_1:\mu \not= 0.
\end{eqnarray*}


\subsubsection{Bayesian model averaged meta-analysis model}
The second hybrid model incorporates the statistical uncertainty regarding the true meta-analysis model via a weighted average of all models under consideration using Bayesian model averaging (BMA), a common approach in Bayesian statistics \citep{hoeting1999bayesian}. A Bayesian model-averaged meta-analysis model \citep{gronau2021primer} can be obtained by averaging over the CE model and the RE model according to
\begin{eqnarray}
\mathcal{M}_{BMA}: y_i \sim p_{CE}\times \mathcal{M}_{CE} + (1-p_{CE}) \times \mathcal{M}_{RE}
\end{eqnarray}
using prespecified prior probabilities for the CE model and RE model, i.e., $\text{Pr}(\mathcal{M}_{CE})=p_{CE}$ and $\text{Pr}(\mathcal{M}_{RE})=1-p_{CE}$. Moreover, each of the two model parts, $\mathcal{M}_{CE}$ and $\mathcal{M}_{RE}$, need to be split regarding the absence and presence of the respective key parameters, i.e., $\theta$ and $\mu$, resulting in four model parts: $\mathcal{M}_{CE} ~\&~ \theta = 0$, $\mathcal{M}_{CE} ~\&~ \theta \not = 0$, $\mathcal{M}_{RE} ~\&~ \mu = 0$, and $\mathcal{M}_{RE} ~\&~ \mu \not= 0$. Typically, equal prior probabilities of $\frac{1}{4}$ are chosen for these four models \citep{gronau2021primer}. Under the BMA approach, the hypothesis test of interest would be formulated as
\begin{eqnarray*}
\mathcal{H}_0&:&(\mathcal{M}_{CE} ~\&~ \theta = 0) \text{ or } (\mathcal{M}_{RE} ~\&~ \mu = 0)\\
\mathcal{H}_1&:&(\mathcal{M}_{CE} ~\&~ \theta \not= 0) \text{ or } (\mathcal{M}_{RE} ~\&~ \mu \not= 0).
\end{eqnarray*}
The BMA approach has also been extended to include (sub)models that correct for publication bias \citep{maier2023robust}. 

\section{Prior specification for the parameters}\label{priors}
%
%
%
%
%
%

Prior distributions (or priors for short) need to be chosen for the parameters under the employed meta-analysis model. Priors reflect the plausibility of the parameter values before observing the data. To test the average effect, proper priors need to be formulated for the average effect under all five models. Additionally, under the RE model, the marema model, and the BMA model, a prior needs to be formulated for the between-study heterogeneity, which is a common nuisance parameter under both $\mathcal{H}_0$ and $\mathcal{H}_1$. Under the RE model and marema model, a noninformative improper prior can be used. Under the BMA model, the prior for the nuisance parameter must be proper (see also Appendix \ref{AppHybrid}).

First we illustrate the sensitivity of Bayes factors to the prior of the tested parameter. Next we discuss prior specification separately for the average effect and for the between-study heterogeneity parameter. Table \ref{defaultpriorstab} gives an overview of default priors which are currently available in existing R packages: BFpack \citep{mulder2021bfpack}, RoBMA \citep{RoBMA}, metaBMA \citep{heck2019}. Because the R package bayesmeta \citep{rover2020bayesian} does not provide default priors for Bayes factor testing, this package is omitted in the table.

\begin{table}[t]
\caption{An overview of available default priors when testing the mean effect using existing R packages: BFpack \citep{mulder2021bfpack}, RoBMA \citep{RoBMA}, and metaBMA \citep{heck2019}.}
\begin{center}
\hspace*{-2cm}
\begin{tabular}{lllllll}
  \hline
& & BFpack & RoBMA & metaBMA \\
\hline
Global effect & SMD & $\mathcal{N}(0,1)$ & $\mathcal{N}(0,1)$ & $\mathcal{C}(0,0.707)$ \\
 & log odds & $t_{13}(0,2.35)$ & Converted $\mathcal{N}(0,1)$ & $\mathcal{C}(0,1.283)$ \\
 & F(correlation) &Logistic$(0.5)$ & Converted $\mathcal{N}(0,1)$ & $\mathcal{C}(0,0.354)$\\
 & general & unit-info. prior\\
Between-study & & Berger-Deely prior & $p(\tau)$=$\mathcal{IG}(1,0.15)$ & $p(\tau)$=$\mathcal{IG}(1,0.15)$\\
heterogeneity & & \\
\hline
\end{tabular}
\end{center}
\label{defaultpriorstab}
\begin{minipage}{0.9\linewidth}
\footnotesize
\textit{Note.} SMD = standardized mean difference; $t_{13}(0,2.35)$ = $t$-distribution with a mean of 0, a scale of 2.35, and 13 degrees of freedom; $\mathcal{C}$ = Cauchy distribution; unit-info. = unit-information; $\mathcal{IG}$ = inverse-gamma distribution.
\end{minipage}
\end{table}

\subsection{Prior sensitivity}
The choice of the prior for the average effect, which is unique under $\mathcal{H}_1$ (as it is fixed under $\mathcal{H}_0$), is particularly important. The sensitivity of the Bayes factor to this prior can be understood from the definition of the Bayes factor in \eqref{BF10}. Under $\mathcal{H}_0$, the average effect is assumed to be fixed at zero, and thus, the marginal probability of the data in the available studies quantifies how likely the data were to be observed under the assumption that the effect is zero. Under $\mathcal{H}_1$, the effect is assumed to be unknown and our belief about its magnitude is reflected in the prior distribution. Thus, the marginal probability of the data is equal to a weighted average of the likelihood of the data weighted according to the specified prior. 

\begin{figure}[htp]
\begin{center}
\includegraphics[height=5.5cm]{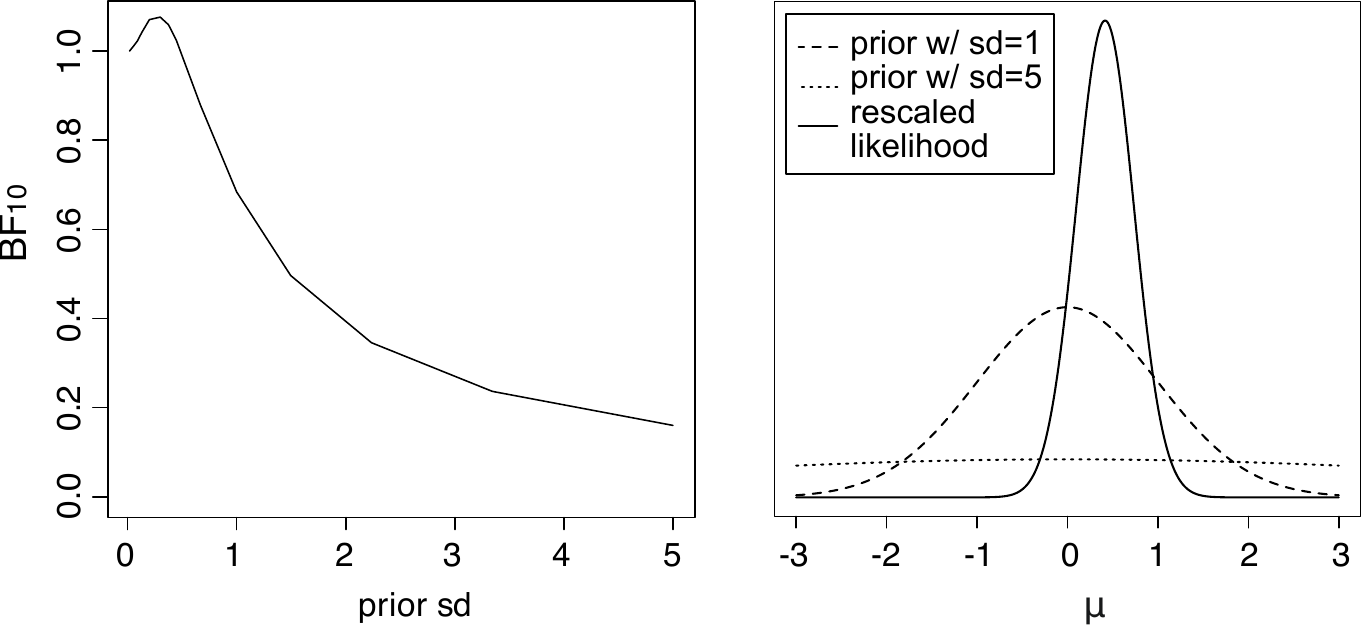}
\caption{Left panel: The Bayes factor $B_{10}$ as a function of the standard deviation of a normal prior for the average effect having a mean of zero for the meta-analysis of \cite{mcneely2010} (Section 2). Right panel: Normal priors with mean 0 and a standard deviation of 1 (dashed line) or 5 (dotted line), and the rescaled likelihood evaluated at $\hat{\tau}=0.243$. The likelihood has its mode at $\hat{\mu}=0.416$.}
\label{priorsens}
\end{center}
\end{figure}

In order for the marginal probability under $\mathcal{H}_1$ to be meaningful, and to ensure that the resulting Bayes factor is meaningful, the prior should correspond to realistic `weights' on the possible (nonzero) effects sizes. For this reason, an extremely vague prior for the average effect should not be used, such as a normal prior with a very large standard deviation. Such a prior would place a relatively large weights at unrealistically large effects sizes, resulting in extremely small marginal likelihoods under $\mathcal{H}_1$ for typical (`small' to `large') effect sizes, thereby heavily biasing the evidence in favor of the null\footnote{This is sometimes referred to as `Bartlett's paradox' \citep[e.g.,][]{Bartlett:1957,Liang:2008}, which was well-understood by \cite{Jeffreys}. Moreover, a noninformative flat improper prior for $\mu$, i.e., $p(\mu)\propto 1$, which is a common default choice in Bayesian estimation, can also not be used in Bayesian testing as the Bayes factor will depend on the arbitrary constant of the improper prior which is only present under $\mathcal{H}_1$.}. This is illustrated in Figure \ref{priorsens} (left panel) when testing the average effect under a RE model for the meta-analysis of \cite{mcneely2010} when placing a normal prior on $\mu$ with a mean of zero and we let the prior standard deviation gradually increase to very large values. As the prior standard deviation of the normal prior increases to unrealistically large effect sizes, the Bayes factor gradually decreases towards zero implying that the evidence in favor of $\mathcal{H}_0$ keeps increasing. Furthermore, the right panel of Figure \ref{priorsens} illustrates that 
when using a prior standard deviation of 5, the prior places lower weights around the likelihood (which is concentrated around $\hat{\mu}=0.416$), than when using a prior standard deviation of 1. Thus, the marginal likelihood of $\mathcal{H}_1$ is lower when a prior standard deviation of 5 is used, as can also be seen from the lower Bayes factor in the left panel in Figure \ref{priorsens}.

%
%
%
%
%
%
%
%
%
%
%
%

\subsection{Priors for the average effect}
The prior for the (average) effect in case of a standardized mean difference can be specified based on different considerations. Because we are testing whether the average effect is zero or not, a natural choice would be to center the prior at zero so that negative effects are equally likely as positive effects, and so that small effects are on average more likely than large effects before observing the data. Moreover, as the distribution of the observed effect given the unknown true effect is also normal, a (conjugate) normal prior would be a natural choice. The choice of the prior standard deviation is particularly important as was illustrated from Bartlett's paradox (Figure \ref{priorsens}). A standard deviation of 1 is the default in the R packages RoBMA \citep{RoBMA} and BFpack \citep{mulder2021bfpack} implying fairly large effects to be plausible. In metaBMA, the more heavy tailed Cauchy prior with scale $0.707$ is the default.

For a log odds ratio as effect size measure (which lies on an approximate normal scale), one could start with placing priors on the success probabilities in the two groups (e.g., treatment and control), which are then transformed to the log odds ratio. As a default choice, proper independent uniform priors can be specified for the success probabilities which assume that all success probabilities are equally likely a priori. By transforming these to the log odds ratio an approximate $t$ distribution having a scale of 2.35 and 13 degrees of freedom is obtained (Appendix B). This is the default in BFpack. Another option could be to start with a prior for a mean effect, e.g., $\mathcal{N}(0,1)$ prior, and convert this prior to the log odds scale the transformation formulas of \citep[][Ch. 7]{borenstein2010}. This is the default in RoBMA and has the advantage that the scale of the prior distribution is approximately the same regardless of whether the standardized mean difference or log odds ratio is the effect size measure of interest. In metaBMA, the default scale of the prior distributions is also adjusted to the effect size measure: a Cauchy prior with scale 1.283 is used, because the distribution of log odds ratio is approximately 1.81 times as wide as of the standardized mean difference. The use of conversion formulas may be particularly useful when the scales of the outcome variables varied across studies in the same meta-analysis. Synthesizing evidence from highly heterogenous outcomes having different measurement levels is only recommended of course when this is substantively meaningful.

Pearson correlation coefficients are commonly meta-analyzed after applying Fisher's $z$ transformation \citep{fisher1915}. Fisher's $z$ transformed correlations are approximately normally distributed. For a Fisher transformed correlation, one can specify a prior for the correlation in the interval $(-1,1)$, which can then be transformed by applying Fisher's $z$ transformation. A natural proper noninformative choice for a correlation would be to use a uniform prior in the interval $(-1,1)$ \citep[see also][]{Jeffreys,mulder2023bayes}. After applying a parameter transformation, this corresponds to a logistic prior distribution with a scale of 0.5 for the Fisher's $z$ transformed correlation (Appendix B). This is the current default in BFpack. Alternatively, one could again use the conversion formulas from \cite{borenstein2010}, which is the default in RoBMA. In metaBMA, a Cauchy prior with a scale of 0.354 is the default to take into account that the distribution of Fisher's $z$ transformed correlations in relation to the distribution of standardized mean differences.

As a general default choice for the prior, that is independent of the effect size measure, a unit-information prior can also be specified which contains the information of a single observation \citep[e.g.,][]{rover2021}. By construction, the amount of prior information is then relative to the amount of information in the sample instead of being based on contextual information about the key parameter. Therefore this prior can be used as a general default. Note that the evidence as quantified by the well-known Bayesian information criterion \citep[BIC;][]{Schwarz:1978} also behaves as an approximate Bayes factor based on a unit-information prior \citep{Raftery:1995,kass1995}. The information of one observation depends on whether between-study heterogeneity is present or not. Under the CE model where between-study heterogeneity is absent, the unit-information prior follows a normal distribution having a mean of 0 and a prior variance that is equal to the error variance rescaled to the total sample size, i.e., $\mathcal{N}(0,N/\sum_i 1/\sigma_i^2)$, where $N=\sum_i n_i$ denotes the total sample size across studies. Under the RE model and the hybrid models, a conditional prior for the average effect $\mu$ is required conditional on $\tau^2$, which follows a normal $\mathcal{N}(0,N/\sum_i(\sigma^2_i+\tau^2)^{-1})$ distribution\footnote{Unlike the unit-information prior of \cite{vanaert2022b} where the `unit' referred to a single study, this prior uses a single observation in a study as the `unit'.}. This prior needs to be combined with a prior for $\tau^2$ (discussed later) to construct a joint prior for $\mu$ and $\tau^2$ under $\mathcal{H}_1$ \citep[Note that such prior dependency between $\mu$ and $\tau^2$ is common in the objective Bayesian literature for Bayes factor testing;][]{Zellner:1986,Liang:2008,bayarri2012criteria}. 

When viewing a prior of the average effect as a population distribution of effect sizes from which the unknown effect size of a current meta-analysis is `drawn', one can use the estimated distribution of effect sizes from published research to create an empirically informed prior. This has been done for meta-analyses of binary data with rare events \citep{gunhan2020random}, for medicine and its subfields \citep{bartovs2021bayesian}, and for binary trial data and time-to-event data \citep{bartovs2023empirical}, for example. Depending on the availability of relevant published effect sizes given the meta-analysis at hand, such an approach may be reasonable. A prior for the average effect has also been elicited for a meta-analysis using external knowledge regarding the effect size at hand \citep{gronau2017bayesian}.

\subsection{Priors for the between-study heterogeneity}\label{SectionPriorHetero}
Under the RE model and the hybrid models, a prior for the between-study heterogeneity also needs to be chosen \citep{rover2020bayesian,Pateras2019,Spiegelhalter2004}. As this is a common nuisance parameter under both hypotheses, $\mathcal{H}_1$ and $\mathcal{H}_0$, the Bayes factor for testing the global effect is considerably less sensitive to the exact choice of this prior \citep[e.g.,][]{Kass:1995}. This feature is advantageous, as specifying an informative prior for the between-study heterogeneity can be challenging due to its less intuitive interpretation. Interestingly, it is also possible to employ a noninformative improper prior for this common nuisance parameter\footnote{\citet{Jeffreys} justified this when the expected Fisher information shows zero cross-information \citep[see also][]{Kass:1992}, which is the case for the RE meta-analysis model. Furthermore, \citet{bayarri2012criteria} showed that in general a Bayes factor based on a proper approximation of an improper prior converges to the Bayes factor based on the improper prior as the proper prior approaches the improper prior in the limit, justifying the use of noninformative improper priors for nuisance parameters.}, allowing a default Bayes factor test.

Currently, the R package `BFpack' supports the use of noninformative improper priors for the between-study heterogeneity. Since the R packages `metaBMA' and `RoBMA' perform Bayesian model averaging between the CE model ($\tau^2 = 0$) and the RE model ($\tau^2 > 0$), calculating Bayes factors is necessary to determine the posterior model weights. This test treats $\tau^2$ as a parameter of interest rather than a nuisance parameter, precluding the use of noninformative improper or arbitrarily vague priors. These two packages do allow a noninformative improper prior for $\tau^2$ when solely working under the RE model (implying that the CE model is disregarded and BMA is not used). Based on our experience, the `bayesmeta' package does not support noninformative improper priors for the nuisance parameter when testing the overall effect.


As summarized by \cite{rover2020bayesian}, there is an extensive literature on noninformative (`objective', `default') priors for the between-study heterogeneity $\tau^2$. For researchers with a less mathematical statistical background, it will be difficult to choose one specific noninformative prior based on the available theoretical and statistical arguments. Moreover, noninformative priors have often been assessed based their implied behavior in Bayesian estimation problems rather than their implied behavior in hypothesis testing using Bayes factors. To keep the discussion as concise as possible, we restrict ourselves to certain noninformative priors rather than providing a complete assessment of all possible priors.

One important criterion for a noninformative improper prior is whether the resulting Bayes factor is well-defined. This is the case when the marginal likelihoods are finite. In an estimation problem, this is equivalent to the important criterion of that the posterior is proper. Table \ref{tab_improppriors} provides three possible noninformative improper priors for the between-study variance when testing the global mean including the minimal number of studies for a well-defined Bayes factor. The uniform prior on $\sqrt{\tau^2}$ or the Berger-Deely prior \citep{berger1988bayesian} may be preferred as Bayes factor testing will already be possible when only two studies are available. From these two priors, the Berger-Deely prior is conjugate under the RE model, and therefore a more natural choice. Moreover, the Berger-Deely prior naturally extends to the marema model\footnote{Under the marema model, the Berger-Deely prior would be $p(\tau^2)\propto\prod_i 1/\sqrt[\leftroot{-2}\uproot{2}k]{\sigma_i^2+\tau^2}$, in the region $\tau^2 > -\sigma_{\min}^2$.} (the Berger-Deely prior is the default in BFpack; Table \ref{defaultpriorstab}). Furthermore, the scale invariant prior, $1/\tau^2$, which is also Jeffreys' prior, is not recommended for lower level variances such as the between-study heterogeneity \citep[e.g., see][]{gelman2006prior}. This prior may result in infinite marginal likelihoods (and improper posteriors in Bayesian estimation) when the between-study heterogeneity in the data is very low, and is therefore not recommended. Finally, we note that proper approximations of noninformative improper priors \citep[e.g.,][]{Pateras2019,Spiegelhalter2001} should be used with care, as they may unduly be highly noninformative \citep{berger2006case,gelman2006prior}.

\begin{table}[t]
\caption{Three possible noninformative priors for the between-study heterogeneity $\tau^2$ and the number of required studies to obtain a finite Bayes factor.}
\begin{center}
\begin{tabular}{lccc}
\hline
 & prior density $p(\tau^2)$ & Number of required studies \\
\hline
uniform prior on $\tau^2$ & 1  & $k\ge 3$ studies\\
uniform prior on $\sqrt{\tau^2}$ & $1/ \sqrt{\tau^2}$ & $k\ge 2$ studies\\
Berger-Deely prior & $\prod_i 1/\sqrt[\leftroot{-2}\uproot{2}k]{\sigma_i^2+\tau^2}$ &  $k\ge 2$ studies\\
\hline
\end{tabular}
\end{center}
\label{tab_improppriors}
\end{table}

Empirically informed priors have also been proposed for the between-study heterogeneity \citep{pullenayegum2011,rhodes2015,turner2015,van2017estimates}. These empirically informed priors have mostly been used for estimation problems, although there is also literature where these priors have also been used for Bayes factor testing in meta-analyses \citep[e.g.,][]{gronau2017bayesian,bartovs2023empirical}. A complicating factor which is often overlooked is that the between-study variance under $\mathcal{H}_1$ will never be larger than the between study variance under $\mathcal{H}_0$ as the mean is restricted under $\mathcal{H}_0$. For this reason, it would be preferred to incorporate this in the informative priors (by ensuring that $\tau^2$ is stochastically larger under $\mathcal{H}_0$ than under $\mathcal{H}_1$ a priori). To our knowledge, no priors have been proposed so far that abide this property. In the end, the suitability of informed priors would need to be carefully assessed depending on the meta-analysis at hand.


\subsection{Final remarks on prior specification}

On a more theoretical note, it has been argued that priors should result in Bayes factors that are information consistent \citep[e.g.,][]{Liang:2008,gronau2020informed,mulder2021prevalence}. Information consistency implies that the evidence for the alternative should go to infinity when the estimated effect goes to plus or minus infinity. Roughly speaking Bayes factors are not information consistent when the (marginal) prior for the mean effect has thinner tails than the (integrated) likelihood as a function of the mean effect (after integrating out the nuisance parameter $\tau^2$). As the integrated likelihood follows an approximate Student $t$ distribution under the RE model, information consistency is assured when using a prior with thicker tails such as a Cauchy prior. As was shown by \cite{mulder2021prevalence}, a normal marginal prior (having very thin tails) can be detrimental when the effect size is extremely large (such as standardized effect sizes of 10) causing the relative evidence to be approximately 1 (suggesting equal evidence for the null and alternative). For the CE and the FE model, the likelihood as function of the mean has a Gaussian shape, which already has thin tails. Therefore, information inconsistency would not even occur when using a normal prior. Based on our experience, information consistency is mainly a theoretical property and not a practical one as effect sizes are generally not that extreme such that an information inconsistent Bayes factor would result in conflicting behavior. Therefore, information consistency may not be a serious concern when choosing priors in general.

Generic approximate Bayesian approaches are also available which avoid manual prior specification, such as the Bayesian information criterion \citep[BIC][]{Schwarz:1978,Raftery:1995} or fractional/intrinsic Bayes factors \citep{Berger:1996,OHagan:1995,Mulder:2014b}, or approximations thereof \citep{Gu:2017}. These methods have for instance been been applied for fixed effects meta-analyses \citep{kuiper2013combining,van2024bayesian,vanLissa2024}. Implicitly, these approximations abide the principle of minimally informative priors, comparable to the unit-information prior. To simplify the interpretation of the evidence however, it may be preferred to only use these approximate methods for hypothesis testing problems which are not supported by the available Bayesian meta-analysis software (e.g., due to the formulated hypotheses, the statistical models, or the research designs).

Finally, Appendix \ref{App_numericalchecks1} presents a small simulation on the sensitivity of the Bayes factor to the prior of the nuisance between-study heterogeneity. As shown, the Bayes factor is quite robust to the exact choice.

\section{Computing Bayes factors for evidence synthesis}\label{compBF}
Depending on the employed meta-analytic model, as well as on the chosen prior (e.g., whether a conjugate prior was used), the complexity of the computation of the Bayes factor differs. Moreover, when a new study becomes available, updating the Bayes factor can be done via different formulas. 

%
%

\subsection{Evidence synthesis via (regular) Bayesian updating}
Under the CE model, the RE model, and the hybrid models, evidence updating is done in a similar manner as regular Bayesian updating in estimation. In Bayesian estimation, we need to update the posterior when a new study is reported. The posterior based on the previous studies becomes the prior, which is then multiplied (`combined') with the likelihood of the new study to obtain the new posterior via Bayes' theorem. When testing hypotheses, we update the Bayes factor based on the previous $k-1$ studies with the Bayes factor for the new $k$-th study using the posteriors based on the previous studies under the hypotheses as prior for computing the marginal likelihoods. Under $\mathcal{H}_1$, this can be written as
\begin{eqnarray}
B_{10}(y_{1:(k+1)}) &=& ~B_{10}(y_{k+1}|y_{1:k})~~\times ~~B_{10}(y_{1:k})\\
&=& \frac{p(y_{k+1}|y_{1:k},\mathcal{H}_1)}{p(y_{k+1}|y_{1:k},\mathcal{H}_0)} \,\times~
\frac{p(y_{1:k}|\mathcal{H}_1)}{p(y_{1:k}|\mathcal{H}_0)}.
\label{update1}
\end{eqnarray}
This follows from basic probability calculus \citep[e.g.,][]{ly2019replication}. Consequently, we can also write the Bayes factor for all $k+1$ studies according to
\begin{eqnarray}
B_{10}(y_{1:(k+1)}) = B_{10}(y_{k+1}|y_{1:k})\times\ldots\times B_{10}(y_{2}|y_{1})\times B_{10}(y_{1}).
\label{update1full}
\end{eqnarray}

Rather than using this updating scheme explicitly, meta-analysts will likely compute the Bayes factor based on the $k+1$ studies ``from scratch'' because R packages generally only have functions for computing marginal likelihoods and Bayes factors for a given set of studies. Computing Bayes factors for a given set of studies is generally done by computing the marginal likelihoods using numerical algorithms (e.g., based on bridge sampling \citep{bennett1976efficient} using the R package `bridgesampling' \citep{gronau2020bridgesampling}, or importance sampling \citep{meng1996simulating}, as used in the R package `BFpack' \citep{mulder2021bfpack}, for instance). When no nuisance parameters are present (as in the CE model) or when the prior of the key parameter is independent of the prior of the nuisance parameter, for instance, it is also possible to compute the Bayes factor using the Savage-Dickey density ratio \citep{Dickey:1971}. This quantity is relatively easy to compute from MCMC output (e.g., using Stan \citep{carpenter2017stan} or JAGS \citep{plummer2003jags}, for example). Here, we briefly explain this as it may give readers some extra intuition since viewing the behavior of Bayes factors as marginal likelihoods may be less intuitive.

\begin{figure}[t]
\begin{center}
\includegraphics[height=6cm]{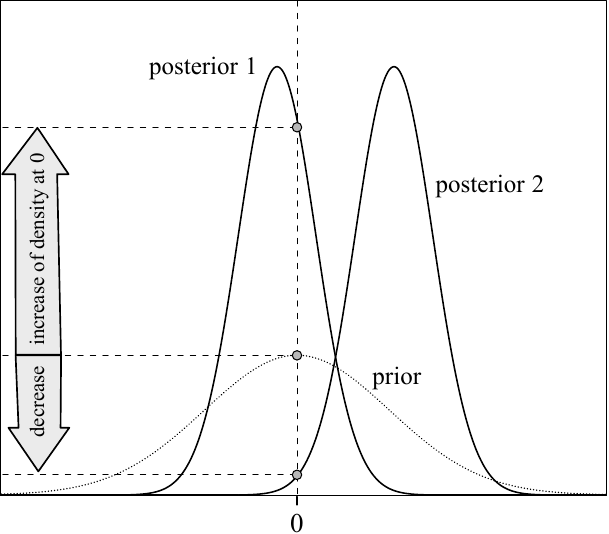}
\caption{Evidence for $\mathcal{H}_0$ based on a Savage-Dickey density ratio. In case of posterior 1, there is evidence for $\mathcal{H}_0$, and in case of posterior 2, there would be evidence for $\mathcal{H}_1$.}
\label{fig_SD}
\end{center}
\end{figure}

The Savage-Dickey density ratio is defined by evaluating the posterior density of $\theta$ under the unconstrained hypothesis $\mathcal{H}_1$, at the null value divided by the unconstrained prior density at the null value \citep{Dickey:1971}:
\[
B_{01}(y_{1:k}) = \frac{p(\theta=0|\mathcal{H}_1,y_{1:k})}
{p(\theta=0|\mathcal{H}_1)}.
\]
Thus, we can compute the Bayes factor in favor of $\mathcal{H}_0$ by simply computing the posterior of $\mu$ (which is obtained from an estimation step) at zero divided by the chosen prior for $\mu$ at zero. The posterior density at the null value can easily be obtained from MCMC output with Bayesian software (e.g., Stan or JAGS). The prior density at the null value generally has an analytic form when the prior belongs to a common family of probability distributions. Consequently, if the density at zero increases from prior to posterior, then there is evidence in favor of the null value, and vice versa (Figure \ref{fig_SD}). In the case a normal prior is considered, the Bayes factor has an analytic expression and thus numerical algorithms can be avoided (Appendix \ref{AppBFcomp}). 

\subsection{Evidence synthesis via the product Bayes factor}
Under the FE model \eqref{fixefmodel}, each study is assumed to have a unique effect size, which are not linked via a distribution on a lower level (as in a RE model). Therefore, the Bayes factor for testing the null hypothesis in \eqref{update1full} can be simplified as the product of the Bayes factors of the separate studies (Appendix \ref{AppBFcomp}):
\begin{equation}
B_{10}(y_{1:k}) = B_{10}(y_{1})\times \ldots \times B_{10}(y_{k}),
\label{productBF}
\end{equation}
which is sometimes called the `product Bayes factor' \citep[e.g.,][]{vanLissa2024}. Consequently, when a new study is reported we can simply multiply the current Bayes factor with the Bayes factor of the new study, and thus, \eqref{update1} can be simplified as
\begin{equation}
B_{10}(y_{1:(k+1)}) = B_{10}(y_{k+1})\times B_{10}(y_{1:k}).
\end{equation}
The Bayes factor for the new study can be computed using the prior for the study specific effect $\theta_k$ rather than requiring the posterior based on the previous studies as in regular Bayesian updating. As no nuisance parameters are present, the individual Bayes factors can also be computed using Savage-Dickey density ratios.

\section{Evidence monitoring and e-value theory}\label{cummeta}

\subsection{Evidence monitoring}


Monitoring the evidence has great potential to minimize research waste by helping researchers, practitioners, patients, and funders to make more informed decisions to start new studies depending on the available statistical evidence regarding the (non)existence of certain effects \citep{lau1992cumulative,chalmers2009avoidable,clarke2014accumulating}. A new study is most likely not initiated if there is, for instance, overwhelming evidence that a treatment is beneficial in a meta-analysis. On the other hand, when the meta-analytic \textit{p}-value is close to the significance threshold or equivalently the null value is very close to the boundary of its confidence interval, a meta-analyst may be motivated to initiate another study to clarify whether the population mean is significantly different from zero (a message that is echoed in Fergusson et al., 2005, for example). Various retrospective meta-analyses showed that many studies were still performed while satisfactory evidence was already available, suggesting inefficient use of existing evidence. This also implies that even a meta-analysis conducted only once can implicitly be cumulative, since many included studies may have been initiated or designed in light of previous results \citep[see also][]{ter2019accumulation}. Together, these observations have led to a call for more evidence-based research \citep{lund2016towards}.

Meta-analysts generally use classical p-value testing to test hypotheses that by default does not take the inherent sequential nature of published studies into account. A consequence of this is that it may result in inflated type I error rates if initiating a new study is based on (repeatedly) testing hypotheses based on evidence of published studies \citep{ter2019accumulation}. As discussed by \cite{higgins2011sequential}, in order to properly control the type I error rate in a classical sequential analysis, restrictive decision rules are necessary including a prespecified maximal amount of information when the sequential analysis need to stop permanently. When this bound is exceeded it is no longer allowed to collect more information and ``it is unclear how to proceed'' \citep{higgins2011sequential}. Still, we cannot claim there is evidence in favor of the null in this case, and to our knowledge no sequential procedures are available for equivalence testing to achieve this. These considerations highlight the complexity of controlling classical error control rates in meta-analyses that get updated. This is particularly problematic as many null hypotheses may in fact be true in applied research \citep{johnson2017reproducibility}.

Using the Bayes factor on the other hand, one can monitor and update the evidence as new studies become available in a flexible manner. Depending on the acquired evidence (e.g., using Figure \ref{fig_BFscale}), a meta-analyst can decide to initiate a new study or not without requiring corrections for multiple testing. The meta-analyst's interpretation of the prior should be taken into account here. If a purely subjective Bayesian approach is considered, which is the case when the Bayes factor is based on informative priors that accurately reflects the meta-analyst's prior beliefs, adding more studies in combination with multiple testing is not a problem (De Heide \& Grunwald, 2021). This dates back to the work on subjective Bayesian statistics (de Finetti (1937), Savage (1972). Loosely speaking, it can be stated Bayes factors generally have good performance (including accurate type I error rates) in the region where the prior is concentrated regardless of the exact rule that is used to decide to start a new study or not (e.g., see also Raftery \& Gill, 2002; Rouder, 2018).

When using noninformative, vague, or default priors, which may be chosen out of convenience (e.g., to avoid careful (potentially time-consuming) specification of informative priors), the story becomes more complicated \citep{hendriksen2021optional,de2021optional}. For example, when using noninformative priors for the between-study heterogeneity when testing the global mean, the Bayes factor may show good performance regardless of the true value or the nuisance parameter. This is because the exact choice of the prior for the nuisance parameter generally does not have a large effect on the Bayes factor. However, recent developments on e-value theory have shown that Bayes factors have good classical error rates, but that this depends on the choice of the prior.

\subsection{Bayes factors and e-value theory}
In the last decade, there has been a considerable development of statistical theory on so-called e-values (`e' for expectation or expected value). For a recent overview of the statistical foundations, see \cite{ramdas2024hypothesis}. This theory provides critical conditions which need to hold in order to control classical type I error rates in sequential data designs in a flexible manner allowing optional stopping/continuation without requiring corrections as in classical p-value testing. Interestingly, there is a close relationship between e-values and Bayes factors \citep[e.g.,][]{grunwald2024safe}. Before discussing this, we first give the general definition of an e-value.

Let $E$ be nonnegative statistical quantity of the relative evidence in favor of an alternative hypothesis against a null hypothesis. Specifically, $E=1$ implies that there is no evidence towards any of the hypotheses, while $E>1$ implies that there is evidence in favor of the alternative $\mathcal{H}_1$. Note that the Bayes factor for $\mathcal{H}_1$ against $\mathcal{H}_0$ also has this interpretation (Figure \ref{fig_BFscale}). Now $E$ is called an e-value, if the expected evidence (or `average evidence') never points towards the alternative when $\mathcal{H}_0$ is true. Mathematically, this can be written as:
\begin{equation}
\label{safetycheck}\mathbb{E}_{\mathcal{H}_0}\{E\} \le 1.
\end{equation}
Thus, rather than restricting the chance of incorrectly falling in the rejection region, as in classical testing, e-value theory restricts the average evidence not pointing towards the wrong hypothesis when $\mathcal{H}_0$ is true. This is a considerably stronger requirement, implying a more conservative test \citep{ramdas2024hypothesis}.


If condition \eqref{safetycheck} holds, it can be shown that the reciprocal of an $E$ value, say, $p_E=1/E$, behaves as a conservative p-value, implying that the probability that $1/E$ is smaller than a prespecified significance threshold, say, $\alpha$, is maximally $\alpha$ under the null, i.e.,
\[
\text{P}_{\mathcal{H}_0}(p_E \le \alpha)\le \alpha.
\]
This is a direct consequence of the definition of an e-value \eqref{safetycheck} using Markov's inequality. Because `$p_E \le \alpha$' is equivalent to `$E \ge 1/\alpha$', we can therefore use the reciprocal of a significance level to construct significance thresholds for $E$ values that ensure that type I error rates are never exceeded (Table \ref{tab_dichot}). As e-values are very conservative, smaller thresholds have also been advocated \citep{ramdas2024hypothesis,shafer1982lindley}. A similar argument was made by \cite{royall2000probability} in the context of likelihood ratios, and by \cite{benjamin2018redefine} for Bayes factors. Furthermore, using theory on super-martingales, we can extend the concept of e-values for a single experiment to sequences of e-values, referred to as e-processes, which ensure evidence monitoring as new studies are reported \citep{ramdas2022testing,shafer2011test,williams1991probability,ville1939etude}. Using e-values for testing, the type I error rate will always be controlled under optional stopping/continuation.

Although the reciprocal of the e-value behaves as a (conservative) p-value, the reverse argument does not hold. The reciprocal of a p-value, i.e., 1/$p$, where a smaller $p$, and thus a larger $1/p$, also implies more evidence against the null, disastrously violates e-value criterion as the expected (average) evidence against the null, while the null is true, is in fact $\infty$ (Appendix \ref{AppSafeBF}).

When no nuisance parameters are present--as in the CE and the FE model--the Bayes factor is always an e-value (the proof is given in Appendix \ref{AppSafeBF}). This is the case regardless of the prior that is specified for the effect(s) under $\mathcal{H}_1$. Now let us assume that after $k$ studies, we observe a meta-analytic Bayes factor of, say, $B_{10}(y_{1:k})=22$ under the CE model. As noted above, we can use the reciprocal of a classical significance level, such as $\alpha=.05$, to construct rejection thresholds for the Bayes factors, say $1/.05=20$ (Table \ref{tab_dichot}). Importantly, this error-based threshold also does not need to be chosen prior to the test, as with classical significance testing \citep{lakens2018justify}, but it can be even chosen post-hoc depending on the observed e-value\footnote{In current statistical practice, researchers often report the smallest significance level that results in a significant result where one, two, or three asterisks refer to a p-value smaller than or equal to .05, .01, or .001, respectively. Clearly, this practice does not control the type I error rate if the significance level is chosen post-hoc.} \citep{ramdas2024hypothesis}. As the observed Bayes factor of 22 exceeds this threshold, we can safely reject the null hypothesis without jeopardizing inflated type I error rates regardless of the intermediate choices that have been made to initiate a new study based on previous outcomes. For a classical test, this is generally not the case. In fact, if nothing is known about the motivations of research groups that initiated the past studies based on findings in earlier studies \citep[implying that we do not know how much of our $\alpha$ was already `spent';][]{ter2019accumulation}, it will be practically impossible to guarantee a type I error rate of maximally .05 if a traditional significance test would have been executed. Using a Bayes factor approach on the other hand, it is even possible to initiate a new study with the aim to obtain more decisive evidence against the null, again without jeopardizing inflated type error rates. This is a consequence of Bayes factors being independent on the sampling plan. It is generally known that p-values depend on the sampling plan \citep[e.g.,][]{Wagenmakers:2007}.


\begin{table}[t]
\caption{Linking thresholds for Bayes factors to common significance levels via $\alpha =1/B_{10}=B_{01}$.}
\begin{center}
\begin{tabular}{lll}
  \hline
threshold for $B_{10}$ & decision \\
\hline
10 & reject $\mathcal{H}_0$ with $\alpha=.10$\\
20 & reject $\mathcal{H}_0$ with $\alpha=.05$\\
100 & reject $\mathcal{H}_0$ with $\alpha=.01$\\
1000 & reject $\mathcal{H}_0$ with $\alpha=.001$\\
\hline
\end{tabular}
\end{center}
\label{tab_dichot}
\end{table}


When nuisance parameters are present--as in RE and hybrid models--the choice of prior for these parameters plays a crucial role in determining whether the e-value criterion is satisfied \citep{hendriksen2021optional,de2021optional}. The reason is that condition \eqref{safetycheck} needs to hold regardless of the true value of the nuisance parameter. As shown by \cite{hendriksen2021optional}, so-called `group invariant' priors are generally recommended. For the between-study heterogeneity however, the noninformative improper scale invariant prior for $\tau^2$ is equal to the Jeffreys prior $1/\tau^2$, which is generally not recommended for lower level variances due to possible infinite marginal likelihoods, as was also explained in Section \ref{SectionPriorHetero}.

To give the reader an indication of whether the noninformative priors in Table \ref{tab_improppriors} and the informative inverse gamma prior result in a Bayes factor that abides the e-value criterion, Appendix \ref{App_numericalchecks2} provides numerical estimates of the expected Bayes factor under the null for different true values of the between-study heterogeneity. These results indicate that a uniform prior on $\tau^2$ may abide the criterion, while the other two noninformative priors only result in very slight violations. The largest violations are observed for the informative $IG(1,0.15)$ prior. Because this informative prior is concentrated around small values (the prior mode of $\tau$ equals $0.075$), the e-value condition would only hold under for small $\tau$ values.

\section{Numerical illustration}\label{comparemodels}
This section aims to give the reader some insights about the different behavior of the evidence quantification under the five different meta-analysis models from Section \ref{SectionMetaModels}. For the illustration, we considered $K=10$ studies, data were generated with effect sizes of 0 (implying that the null is true), and standard error were generated from uniform distributions in the interval $(.2,.8)$. To see how the evidence depends on the between-study heterogeneity, we varied $\tau$ on a grid from 0 to 1. The analyses were done using R \citep{rcoreteam2024} using the R package `BFpack' \citep{mulder2021bfpack} except for the BMA results which were obtained using the R package metaBMA \citep{heck2019}. Equal prior probabilities were used for the hypotheses in the BMA method.

\begin{figure}[t]
\begin{center}
\includegraphics[width=\textwidth]{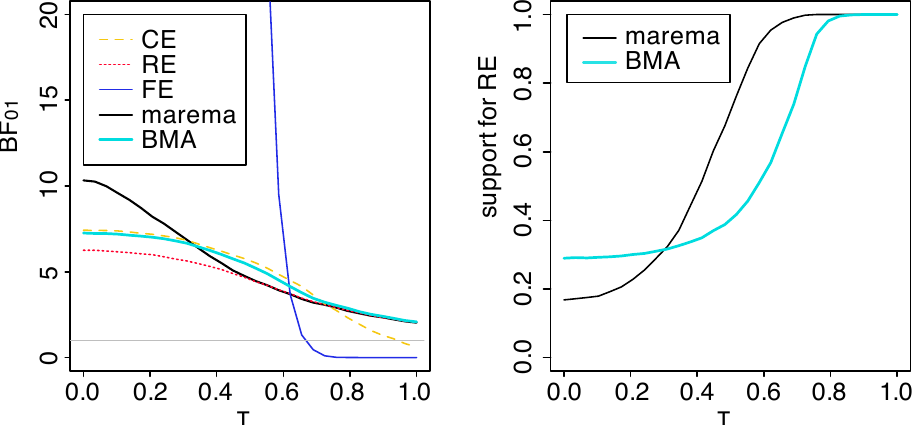}
\caption{The evidence for $\mathcal{H}_0$ against $\mathcal{H}_1$ (left panel) and the posterior support for a RE model under the two hybrid models (right panel) as a function of the between-study variation.}
\label{fig_illustration}
\end{center}
\end{figure}

Figure \ref{fig_illustration} shows the median of the evidence for $\mathcal{H}_0$ against $\mathcal{H}_1$ (left panel). The FE model shows the most striking difference with the other four models. This is a consequence of a different null hypothesis that is tested potentially resulting in extreme evidence quantifications. For example for moderate between-study heterogeneity ($\tau\approx 0.75$), the FE model results in very strong evidence against the null while the other approaches result in mild evidence in favor of the null. Furthermore, we see that in the case of a large amount of between-study heterogeneity, the RE, marema, and BMA models behaves virtually identical. When the between-study heterogeneity is smaller, the behavior of the BMA model `switched' relatively fast towards the CE model, while the marema model behoves more similar to the RE for a longer time. When the between-study heterogeneity is very small or zero, the evidence for the null is lowest for the RE model and largest for the marema model (if we ignore the FE model in this comparison as it tests a completely different null). Moreover, the right panel of Figure \ref{fig_illustration} shows the median support for the existence of random effects, which is quantified as the posterior probability of $\tau^2>0$ under the marema model, while the BMA model uses the posterior model probability of the RE model (which requires prior model probabilities and a proper prior for $\tau^2$ under the RE model; Appendix A). The plot shows a comparable trend of the two methods. The support for nonexistence or existence of random effects is more pronounced under the marema model in comparison to the BMA approach.

\section{Bayesian evidence synthesis in two empirical meta-analyses} 
This section describes the application of the five Bayesian evidence synthesis methods in two meta-analysis examples. We computed the Bayes factors and posterior model probabilities using the implementation in the R package \texttt{BFpack} \citep{mulder2021bfpack} for the CE, RE, and the marema model. \texttt{BFpack} was also used for computing Bayes factors of the individual studies that are used as input for the Bayesian evidence synthesis under the FE model. The R package \texttt{metaBMA} \citep{heck2019} was used for analyzing the data with the BMA model. Results of the frequentist CE and RE model are also reported for comparison. These results were obtained using the R package \texttt{metafor} \citep{viechtbauer2010}. The R codes of the analyses are available at https://osf.io/8h5n6/files/rdp3c and https://osf.io/8h5n6/files/5b48r.

\subsection{Example 1: Statistical learning of people with language impairment}

\begin{figure}[t]
\begin{center}
\includegraphics[width=\textwidth]{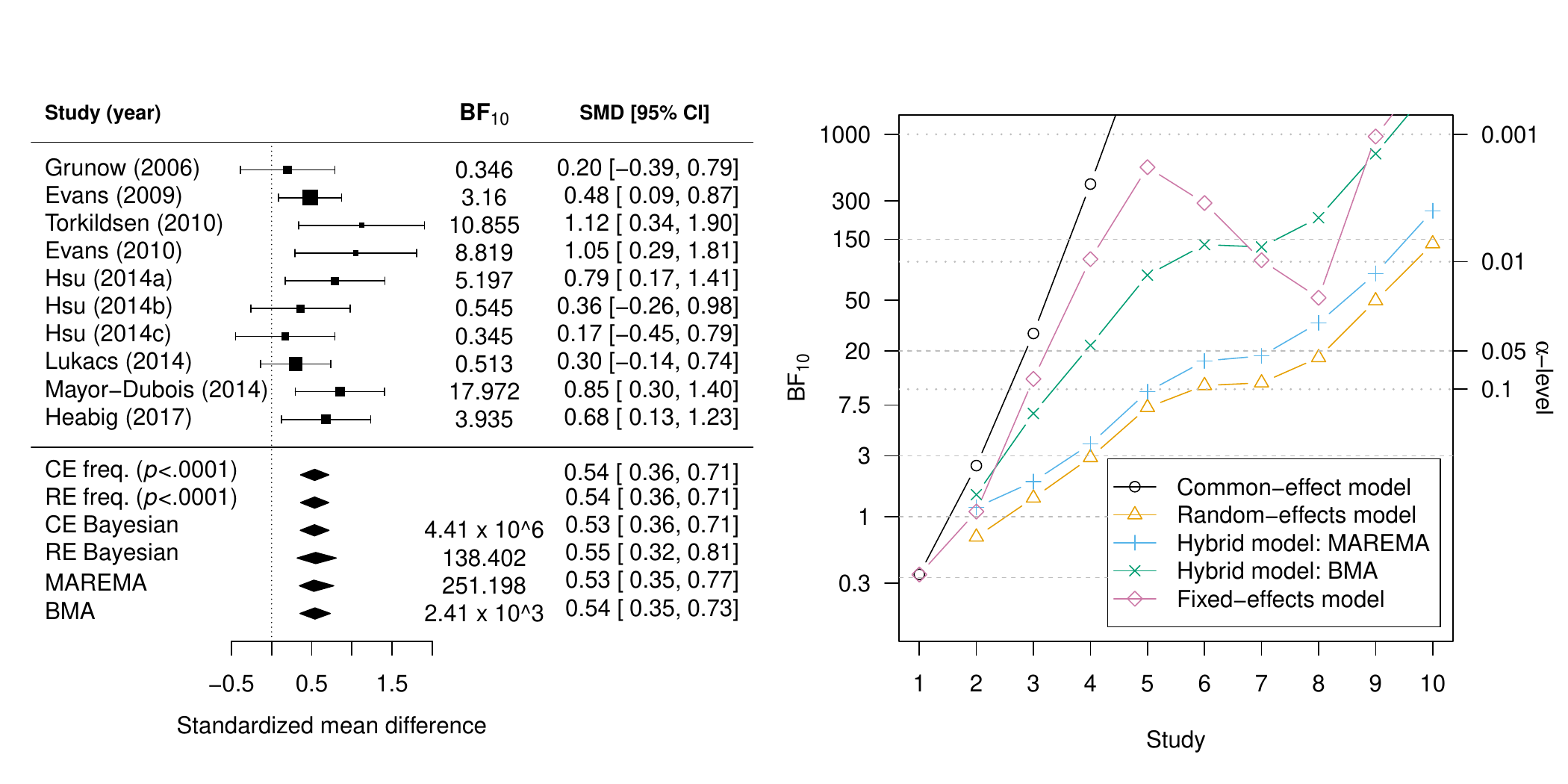}
\caption{Forest plot and the results of Bayesian updating for the meta-analysis of Lammertink et al. (2017). Estimates of the mean effect size of the Bayesian methods were obtained using a normal prior with mean zero and standard deviation of 1,000 for the mean effect.}
\label{fig_lammertink}
\end{center}
\end{figure}

The first example is a meta-analysis presented by \cite{lammertink2017} on the difference in sequential statistical learning ability between people with and without specific language impairment. Sequential statistical learning refers to the ability to learn structures in text by, for instance, listening to people having a conversation. The goal of the meta-analysis was to assess whether people with a language impairment scored differently on statistical learning than people without such an impairment. Ten effect sizes were included in the meta-analysis. Hedges' \textit{g} standardized mean differences were reported for each study where a larger Hedges' \textit{g} is indicative that people without a language impairment outperformed those with an impairment. Data of this meta-analysis are presented in a forest plot in the left panel of Figure \ref{fig_lammertink}.

We used the proposed normal prior with mean 0 and standard deviation 1 for the mean effect since standardized mean difference was the effect size measure in this meta-analysis. The Berger-Deely prior was used for the between-study heterogeneity in the RE and marema models. For metaBMA, the prior $\mathcal{IG}(1,0.15)$ was used for $\tau$ since this is the default in this R package and metaBMA does not allow noninformative improper priors. 

\cite{lammertink2017} opted for a RE model even though $\tau^2$ was estimated to be zero and the null hypothesis of no between-study variance was not rejected with the $Q$-test (\textit{Q}(9)=10.126, \textit{p}=.340) in the frequentist test. This indicated that there was insufficient evidence to reject a CE model against a RE model. Also the hybrid models showed no strong evidence for the presence of between-study heterogeneity with a posterior probability of 0.570 for $\tau^2 > 0$ under the marema model, and the BMA model resulted in a posterior probability of 0.355 for a RE model.

The first two columns of Table \ref{tab:results_apps} show the Bayes factors of $\mathcal{H}_1$ vs. $\mathcal{H}_0$ and posterior probabilities for $\mathcal{H}_1$. Based on all 10 studies, all methods provided very strong to extreme support for $\mathcal{H}_1$ as all Bayes factors were large and the posterior probabilities of $\mathcal{H}_1$ were close to one. Hence, there is convincing evidence to conclude that people with a language impairment score differently on statistical learning than people without such an impairment. Note here that the posterior probabilities can be also be viewed as conditional error probabilities \citep[e.g.,][]{hoijtink2019tutorial}. For example, if one would conclude that $\mathcal{H}_1$ is true under the RE model, there would be a probability of 0.007 to draw the wrong conclusion given the available data. 

The right panel of Figure \ref{fig_lammertink} shows the Bayes factors using a sequential updating approach where a study's publication year determined the order of the studies. The horizontal gray dashed lines indicate the different categories of evidence for the Bayes factor (Figure \ref{fig_BFscale}) and the dotted lines show the common significance thresholds (Table \ref{tab_dichot}). These results show that $\mathcal{H}_0$ could `safely' be rejected at $\alpha = 0.001$ under the CE model and BMA after these 10 studies regardless of the decisions were made to start each of these studies. For the RE and marema model, the evidence exceeds the threshold $\alpha = 0.01$ after 10 studies. Also note in the right panel of Figure \ref{fig_lammertink} that under the FE model the evidence for $\mathcal{H}_1$ decreased at some point. This can be explained by obtaining evidence in favor of the null in studies 6, 7 and 8, illustrating that the evidence under the FE model can be highly sensitive to the estimated effects of individual studies.


We also ran several sensitivity analyses to examine whether the results are robust to the prior of the nuisance parameter. These sensitivity analyses showed that the results hardly changed when different prior distributions were used. Details of these sensitivity analyses are reported in Appendix \ref{AppEmpEx}.


\begin{table}[]
\caption{Bayes factors (BF$_{10}$) and posterior probabilities ($PHP(\mathcal{H}_1)$) for testing $\mathcal{H}_1$ versus $\mathcal{H}_0$ based on all available studies. }
\label{tab:results_apps}
\begin{tabular}{lcccc}
\hline
                     & \multicolumn{2}{c}{Lammertink et al. (2017)} & \multicolumn{2}{c}{McNeely et al. (2010)} \\ \hline
                     & $BF_{10}$              & $PHP(\mathcal{H}_1)$          & $BF_{10}$           & $PHP(\mathcal{H}_1)$          \\ \cline{2-5} 
CE model  & $4.41 \times 10^6$          & 1                   & 1.112               & 0.526               \\
RE model & 138.402                & 0.993               & 0.324               & 0.245               \\
Hybrid model: marema & 251.198                & 0.996               & 0.349               & 0.259               \\
Hybrid model: BMA    & $2.41 \times 10^3$            & 1                   & 0.659               & 0.397               \\
FE model  & $3.71 \times 10^3$              & 1                   & 0.195               & 0.163               \\ \hline
\end{tabular}
\end{table}

\subsection{Example 2: Exercising after a breast cancer surgery}
\begin{figure}[thp]
\begin{center}
\includegraphics[width=\textwidth]{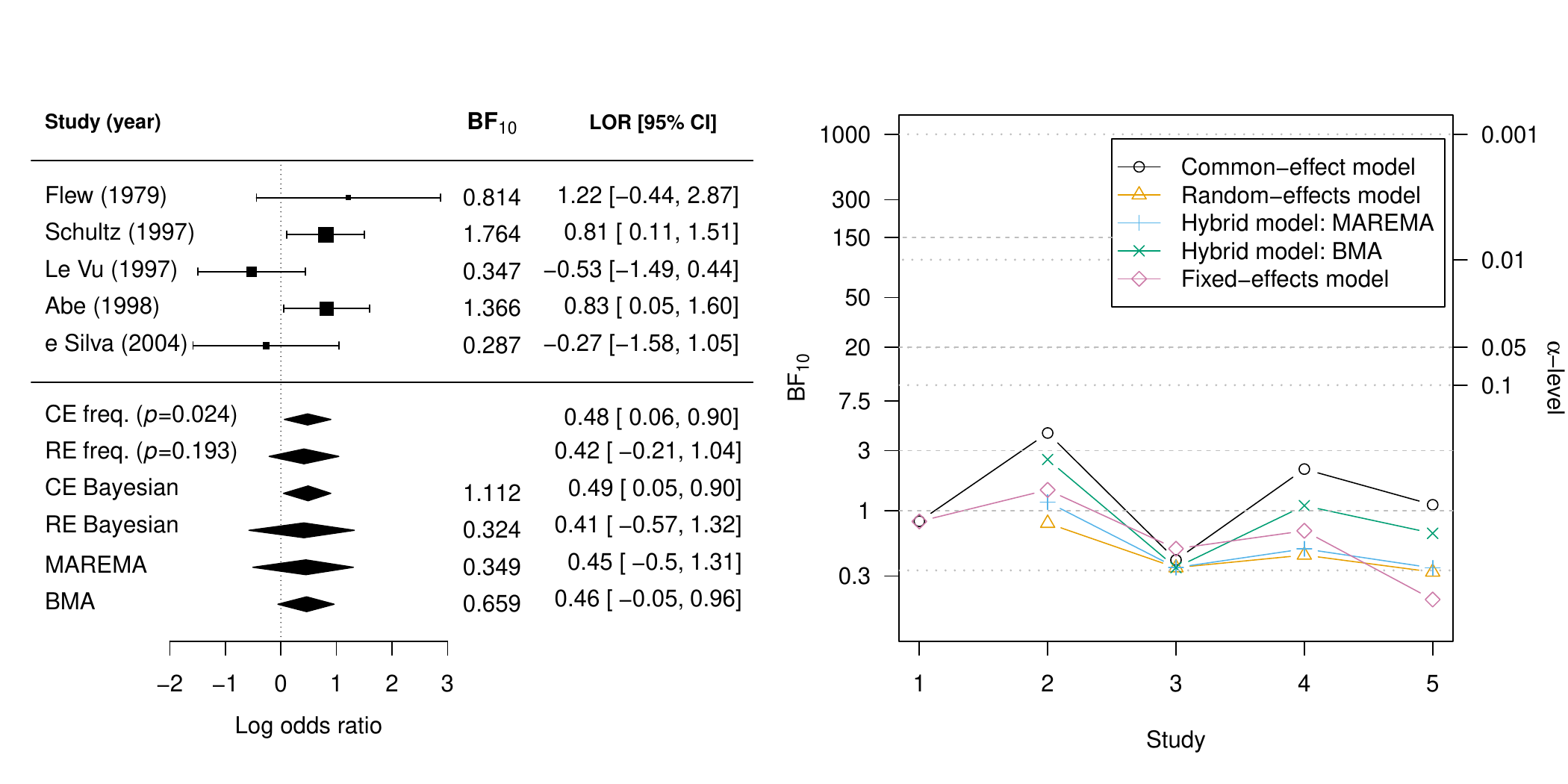}
\caption{Forest plot and the results of Bayesian updating for the meta-analysis of McNeely et al. (2010). Estimates of the mean effect size of the Bayesian methods were obtained using a normal prior with mean zero and standard deviation of 1,000 for the mean effect.}
\label{fig_mcneely1}
\end{center}
\end{figure}

The second example is a meta-analysis by \cite{mcneely2010} on the incidence of seroma when patients start exercising within or after three days following a breast cancer surgery. Five studies are included in this meta-analysis where patients were assigned to an early or delayed exercise condition in each study. The outcome variable was the occurrence of seroma. Thus, a log odds ratio was the effect size measure of interest. A log odds ratio larger (smaller) than one indicates that seroma is more (less) likely to appear in this early period compared to delayed exercise condition. The data of this meta-analysis are presented in the forest plot in the left panel of Figure \ref{fig_mcneely1}.

Using uniform priors on the occurrence of seroma in both conditions implies an approximate Student $t$ prior with a scale of 2.35 and 13 degrees of freedom for log odds ratio. Again, the Berger-Deely prior was used for the between-study heterogeneity in the random effects and marema models, and the empirically informed inverse-gamma prior was used in the BMA model. \cite{mcneely2010} used a RE model in their meta-analysis even though the null hypothesis of no between-study heterogeneity was not rejected with the $Q$-test (\textit{Q}(4)=7.765, \textit{p}=.101). This test was likely to be underpowered since there were only five studies included in this meta-analysis. The posterior probabilities of $\tau^2 > 0$ under the marema model was equal to 0.918, implying considerable evidence for positive between-study variance. Under the BMA model, a more conservative outcome was obtained yielding a posterior probability for the RE model of 0.649. Note that the outcome under the marema model gives a more rigid indication of the support for random effects while the BMA model behaves a smoother (Section \ref{comparemodels}).

The Bayes factors and posterior probabilities for $\mathcal{H}_1$ under the different models are reported in the last two columns of Table \ref{tab:results_apps}. The results show inconclusive results with only mild evidence in the direction of $\mathcal{H}_0$, except under the CE model (which most likely was incorrectly specified given the support for $\tau>0$). Moreover, the right panel in Figure \ref{fig_mcneely1} shows that the differences between the different models were small in the sequential updating approach. These results were in line with those of the frequentist meta-analysis (last two rows for the forest plot in Figure \ref{fig_mcneely1}) where the null hypothesis of no average effect was rejected under the CE model ($z=2.259$, $p=.024$) but not under the RE model ($z=1.300$, $p=.194$). Due to the absence of evidence, more studies are needed to draw a more reliable conclusion about the effect of exercising on the incidence of seroma. Importantly, the type I error rate would not be at stake when using the Bayes factor as test statistic in follow-up tests of this meta-analysis. For a classical test, this would have been problematic however. For example, under the RE model, a nonsignificant effect is obtained based on a significance level of .05, implying that the complete type I error probability of .05 has already been `spent' \citep{ter2019accumulation}, and therefore it would be `unclear how to proceed' \citep{higgins2011sequential}.

We also studied the extent to which the results were sensitive to using different prior distributions for the nuisance parameter. These sensitivity analyses showed that the results hardly changed when different prior distributions were used. Details about these sensitivity analyses are reported in Appendix \ref{AppEmpEx}.

\section{Discussion}

This paper highlighted the practical benefits and the methodological considerations when performing a Bayes factor test in a meta-analysis. The test was discussed for five different meta-analysis models. Table \ref{tabBES} gives an overview of the five approaches, the key assumptions, the hypothesis test, and the formula for Bayesian updating. This overview aims to guide researchers to apply the appropriate evidence synthesis model depending on the meta-analysis at hand.

\begin{table}[t]
\caption{Overview of models and applications for Bayesian evidence synthesis.}
\begin{tabular}{lcccccccccc}
  \hline
model 					& assumption 		& Hypotheses on 	& Evidence updating  \\
\hline
Common effect		& common effect & common effect ($\theta$)   	& Sequential Bayes & \\
model		& across studies	&  		& factors \eqref{update1full} &\\
\\
Random effects		& different effects & average effect ($\mu$) 	& Sequential Bayes \\
model				& across studies	& 		& factors \eqref{update1full} \\
\\
Hybrid models:		& common/different  & average effect ($\mu$) & Sequential Bayes & \\
marema, BMA				& effects across &  & factors \eqref{update1full} & \\
					& studies			&     & &  \\
\\
Fixed effects		& different effects & all study-specific   & Product Bayes 	& \\
model		& across studies	& effects ($\theta_1,\ldots,\theta_k$)		& factors \eqref{productBF} & \\
\hline
\end{tabular}
\label{tabBES}
\end{table}

Because of the importance of the prior, in particular for the global effect under the alternative hypothesis, prior specification was thoroughly discussed. Depending on the nature of the data (log odds, standardized mean difference, or correlations), different choices may be considered. For the between-study heterogeneity, a common nuisance parameter under both $\mathcal{H}_0$ and $\mathcal{H}_1$, noninformative priors were discussed as well as an informative inverse gamma prior. Although noninformative improper priors have been largely overlooked when computing Bayes factors in meta-analyses, noninformative priors are useful for a default Bayes factor test.

Of particular interest may be the hybrid effects models (the marema and the BMA approach), which avoid the error-prone dichotomous decision to choose between the CE and RE model. These hybrid models automatically behave according to a CE or RE formulation depending on the amount of between-study heterogeneity. Moreover, they provide Bayesian quantifications for the support for a random effects formulation without relying on large-sample theory such as the commonly used $Q$-test.

It was also shown how the FE model fundamentally differs from the other approaches in the null hypothesis that is tested. Under the FE model, all study-specific means are tested to be zero, which can be implemented by taking the product of the Bayes factors from the separate studies, whereas the other approaches focus on testing a global effect and treat the individual study effects as nuisance parameters. As noted by \citet{rouder2011bayes} in their response to \cite{wagenmakers2011psychologists} and \cite{bem2011feeling}, aggregating evidence across studies by multiplying study-specific Bayes factors can yield incoherent inference when the goal is to test a common, global effect. This incoherence arises because each study is evaluated under its own parameterization, which is not linked to a shared global effect. For example, when individual studies based on small samples all yield small effect sizes, we may observe weak evidence in favor of the null for all separate studies, and thus multiplying these Bayes factors would produce strong support for the null. In contrast, a joint analysis either using a CE, RE, or hybrid model could yield strong evidence for a nonzero global effect due to the increased precision because of the  pooled data. Conceptually, similar contradictions may emerge using other approaches. For instance, separate confidence intervals of the study-specific effects may all include zero, while the confidence interval for the global effect (under either a FE or CE model) may exclude zero. Therefore, the product-based FE approach should not be used when the interest lies in testing a global effect, which is typically the main objective in meta-analytic applications. The FE model may still be useful when the study-specific effects are defined on different scales (and the interest is also in testing all separate effects jointly), although it should then be carefully assessed whether combining the evidence is substantively meaningful given the large between-study heterogeneity.

Finally, due to the recent developments of e-value theory, it is known that Bayes factors can be transformed to conservative p-values while at the same time they are not sensitive to the (generally unknown) past decisions to start new studies (such as the decision to initiate a new study because of `significant' findings in previous studies). Therefore, Bayes factors can be used for significance-based testing without risking inflated type I errors (unlike classical p-values). This robustness is particularly relevant because even seemingly non-sequential meta-analyses often have an implicit cumulative nature. This makes the Bayes factor a highly flexible statistical testing procedure for meta-analyses where the evidence is currently inconclusive.


\section*{Author contributions}
Joris Mulder: Conceptualization; methodology; software; writing. Robbie van Aert: Conceptualization; methodology; writing.

\section*{Data Availability}
Code can be found here: \url{https://osf.io/8h5n6}

\section*{Funding}
None.

\section*{Competing interests}
None.

\bibliographystyle{apacite}
\bibliography{refs_mulder}

\appendix

\section{Differences between hybrid meta-analytic models}\label{AppHybrid}
Although both the marema model and the BMA meta-analytic model incorporate model uncertainty regarding the true state of the heterogeneity across studies, there are important differences in both approaches. First, the BMA approach requires the manual specification of prior model probabilities across the four modeling parts. Although equal prior weights may be a reasonable default choice other (reasonable) choices can also be made. For example, in the case of a meta-analysis with many reported studies (i.e., $k$ is large), the assumption of study homogeneity may be less likely implying that a smaller prior weight for the CE model may be more realistic. The question is then how large (or small) should these prior probabilities be? For the marema model, no manual specification is required because the sign of the between-study heterogeneity $\tau^2$, which is a natural part of the likelihood, defines the distinction between the common effect and random effects model. Therefore, as the number of studies $k$ grows, the smallest sample variance, $\sigma_{\min}^2$, is likely to decrease inducing a smaller region where $\tau^2$ is negative. 

A second difference is the choice of the prior for the free parameters. As mentioned earlier, the marema model allows noninformative improper priors for the between-study heterogeneity when testing the global model as well as for testing for the presence of between-study heterogeneity (via $\tau^2>0$). Under the BMA approach, a proper (meaningful) prior for $\tau^2$ is required. Depending on the application, this may be a complicating factor.


A third difference we note here is that the BMA meta-analytic model can easily be extended to include other possible (sub)models, such as selection models that take into account publication bias, a persistent problem in meta-analysis \citep{maier2023robust}. Of course, it would also be possible to extend the marema model using BMA methodology to incorporate models that correct for publication bias.

\section{Mathematical details on the prior}
\begin{figure}[t]
\begin{center}
\includegraphics[width=9cm]{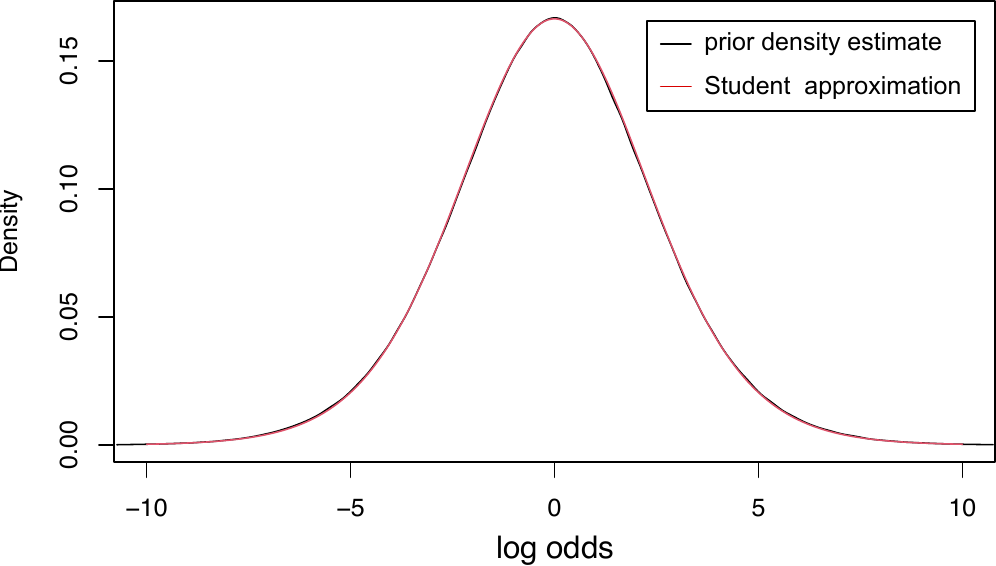}
\caption{Density estimate of the default log odds based on uniform priors for the success probabilities (black line) and Student $t$ approximation with scale 2.35 and 13 degrees of freedom.}
\label{fig_prior_logodds}
\end{center}
\end{figure}

To obtain the default prior for the log odds, we start with uniform priors for the success probabilities under the two conditions, i.e., $p_1 \sim U(0,1)$ and $p_2 \sim U(0,1)$. The log odds is defined by $\theta=\log \left(\frac{p_1}{1-p_1}\big{/}\frac{p_2}{1-p_2}\right)$. Although results are available for the distribution of the risk, i.e., $\frac{p_1}{p_2}$ \citep{pham2000distributions}, to our knowledge no analytic results are available for the (log) odds. Therefore, a numerical approximation is considered. Figure \ref{fig_prior_logodds} shows the estimated density estimate (black line) and the Student $t$ approximation with scale 2.35 and 13 degrees of freedom (red line) which was estimated using the \texttt{fit.st()} function from the \texttt{QRM} package \citep{pfaff2016package}. The analytic approximation is virtually indistinguishable from the numerical estimate of the true prior distribution.

To obtain the default prior for the correlation, $\eta$, we start with a uniform prior in the interval $(-1,1)$ having density $p(\rho)=0.5$. The Fisher's $z$ transformed correlation can be written as $\eta=\frac{1}{2}\log\left(\frac{1+\rho}{1-\rho}\right)$. Consequently, $\rho=\frac{\exp\{2\eta\}-1}{\exp\{2\eta\}+1}$. The Jacobian of the transformation is thus given by $\frac{d\rho}{d\eta}=\frac{4\exp\{2\eta\}}{(1+\exp\{2\eta\})^2}$, and the prior of Fisher's $z$ transformed correlation is thus, $p(\eta)=\frac{2\exp\{2\eta\}}{(1+\exp\{2\eta\})^2}$, which corresponds to a logistic distribution with scale 0.5.


\section{Robustness of the Bayes factor to the prior of the nuisance parameter}\label{App_numericalchecks1}
A numerical simulation was done to assess the sensitivity of the Bayes factor for testing the global mean in a random effects meta-analysis model when using different priors for the (nuisance) between-study heterogeneity. The following conditions were included: 
\begin{itemize}
\item Number of studies: $k=3$, 8 or 20.
\item The global mean: $\mu=0$, .2, .5, or 1.
\item Between-study standard deviation: $\tau=.1,$ .5, or 2.
\item Priors for the between-study heterogeneity: 
\begin{itemize}
\item Improper uniform prior on $\tau^2$: $p(\tau^2)=1$.
\item Improper uniform prior on $\tau$: $p(\tau^2)=1/\sqrt{\tau^2}$.
\item Improper Berger-Deely prior \citep{berger1988bayesian}.
\item Proper inverse gamma prior: $\tau\sim IG(1,0.15)$.
\end{itemize}
\item Study-specific error standard errors $\sigma_i$ were sampled from a uniform distribution: $\mathcal{U}(.2,.8)$.
\end{itemize}
The prior for the global effect $\mu$ was set to a standard normal distribution, the default for a standardized mean difference in RoBMA \citep{RoBMA} and BFpack \citep{mulder2021bfpack}. Under each condition 2,000 data sets were generated and Bayes factors were computed for the two-sided test of the global effect under the random effects meta-analysis model.

\begin{figure}[htp]
\hspace{-2cm}
\includegraphics[width=15cm]{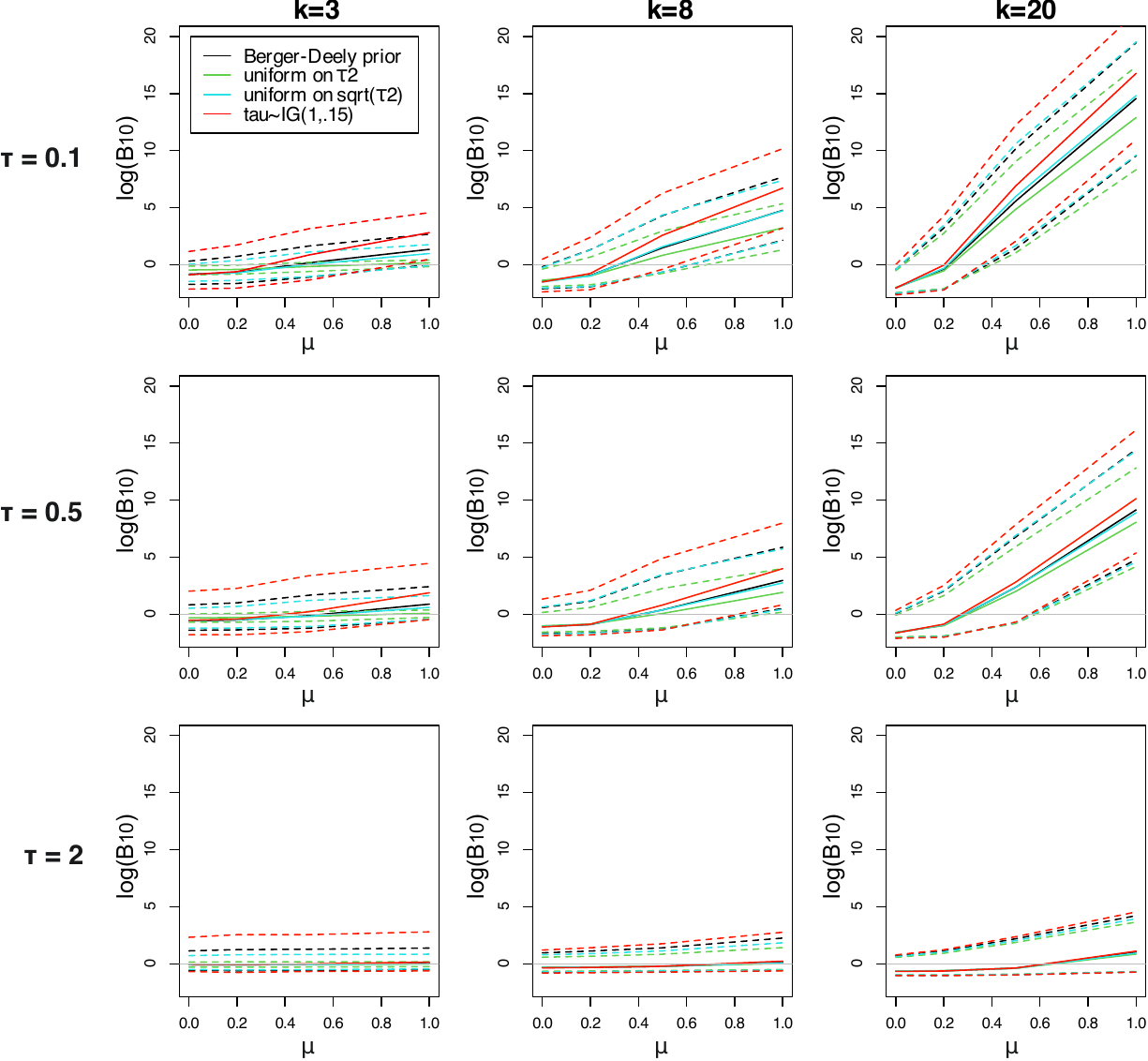}
\caption{5\%, 50\% (solid lines), and 95\% quantiles of the sampling distribution of the Bayes factor using on different priors for the nuisance parameters $\tau^2$ based on 2,000 randomly generated data sets.}
\label{fig_priorsens}
\end{figure}

Figure \ref{fig_priorsens} shows the 5\%, 50\% (median), and 95\% quantiles of the sampling distributions of the logarithm of the Bayes factor as function of the true global mean. First, we see that the sampling distributions of the Bayes factor are hardly affected by the exact choice of the prior of the between-study heterogeneity. Moreover, the plots show the anticipated behavior where we generally obtain evidence for the null if the null is true and the evidence for the alternative hypotheses increases for larger effect sizes. The evidence also becomes more pronounced when the number of studies is large and when the degree of between-study heterogeneity is relatively low. This behavior can be explained by the fact that there is more information available in the data in these cases.

\section{Derivations for the Bayes factor}\label{AppBFcomp}
\subsection{Analytic expression of the Bayes factor under the common effect model with normal prior}
When using a conjugate normal prior for the mean effect, $\theta\sim\mathcal{N}(0,\sigma_0^2)$, under the alternative common effect model, the posterior based on $k$ studies also follows a normal distribution with mean and variance given by
\begin{eqnarray*}
m_k =\frac{\sum_i y_i /\sigma_i^{2}}{1/\sigma_0^2+\sum_i 1/\sigma_i^{2}}\text{~ and~ }
v^2_k=\frac{1}{1/\sigma_0^{2}+\sum_i 1/\sigma_i^{2}}.
\end{eqnarray*}
and the Bayes factor would have a simple analytic form
\begin{equation}
B_{01}(y_{1:k}) = \frac{p(\theta=0|y_{1:k},\mathcal{H}_1)}{p(\theta=0|\mathcal{H}_1)}=\frac{N(0|m_k,v_k^2)}{N(0|0,\sigma_0^2)} = \frac{\sigma_0}{v_k}\exp\left\{-
\frac{m_k^2}{2v_k^2}
\right\}.
\label{BFa}
\end{equation}
In the general case of manually specified priors, the posterior would not belong to a known probability distribution, and the Bayes factor would not have an analytic expression. In that case numerical techniques are required for the computation.

Due to the simple analytical form of the Bayes factor, the evidence can easily be updated when data from a new study $k+1$ become available. One only needs to update the posterior mean and the posterior variance with the new estimate, $y_{k+1}$, and variance, $\sigma_k^2$, according to
\begin{eqnarray*}
m_{k+1} = \frac{y_{k+1}/\sigma_{k+1}^2 + m_k/v^2_k}{1/\sigma_{k+1}^2 + 1/v^2_k}\text{~ and~ }
v^2_{k+1} = \frac{1}{1/\sigma_{k+1}^2 + 1/v^2_k}.
\end{eqnarray*}
The synthesized Bayes factor \eqref{BFa} can then be computed from these updated posterior quantities.

\subsection{Derivation of the product Bayes factor under the fixed effects model}
Here we prove that the Bayes factor for the null hypothesis, $\mathcal{H}_0:\theta_1=\ldots=\theta_k=0$, against the alternative $\mathcal{H}_1:~\text{not }\mathcal{H}_0$, under the fixed effects model is equal to the product of the Bayes factors of every separate effect being zero or not when using independent priors for the study-specific effect sizes under $\mathcal{H}_1$. The proof follows standard probability calculus:
\begin{eqnarray*}
B_{01} &=& \frac{p(y_{1:k}|\mathcal{H}_0)}{p(y_{1:k}|\mathcal{H}_1)}\\
&=& \frac{p(y_{1}|\theta_1=0)\cdots p(y_{k}|\theta_k=0)}
{\int\cdots\int p(y_{1}|\theta_1)\cdots  p(y_{k}|\theta_k) p(\theta_1|\mathcal{H}_1)\cdots p(\theta_k|\mathcal{H}_1)d\theta_1\cdots d\theta_k}\\
&=& \frac{p(y_{1}|\theta_1=0)\cdots p(y_{k}|\theta_k=0)}
{\int p(y_{1}|\theta_1)p(\theta_1|\mathcal{H}_1)d\theta_1\cdots \int p(y_{k}|\theta_k)p(\theta_k|\mathcal{H}_1)d\theta_k}\\
&=& \frac{p(y_{1}|\theta_1=0)}{\int p(y_{1}|\theta_1)p(\theta_1|\mathcal{H}_1)d\theta_1}
\cdots \frac{p(y_{k}|\theta_k=0)}{\int p(y_{k}|\theta_k)p(\theta_k|\mathcal{H}_1)d\theta_k},
\end{eqnarray*}
where $\frac{p(y_{i}|\theta_i=0)}{\int p(y_{i}|\theta_i)p(\theta_i|\mathcal{H}_1)d\theta_i}$ is equal to the Bayes factor of $\mathcal{H}_{0,i}:\theta_i=0$ versus $\mathcal{H}_{1,i}:\theta_i\not =0$, which completes the proof.

\section{Proofs of e-value criterion under the common effect and fixed effects model}\label{AppSafeBF}
Although the reciprocal of an e-value, say, $1/E$, behaves as a (conservative) p-value, the reverse argument is severely violated. Let the p-value be denoted by $p$. The expected (average) evidence as quantified by the reciprocal of the p-value is then given by:
\[
\mathbb{E}_{\mathcal{H}_0}\{1/p\}=\int_{0}^1 1/p ~ dp=\infty.
\]

Next we show that the Bayes factor $B_{10}$ satisfies the e-value condition \eqref{safetycheck} under the common effect and fixed effects models (which do not contain any nuisance parameters), implying that the Bayes factor is an e-value. The proof only requires basic probability calculus:
\begin{eqnarray}
\nonumber\mathbb{E}_{\mathcal{H}_0}\{B_{10}(y_{1:k})\} &=& \int B_{10}(y_{1:k})p(y_{1:k}|\mathcal{H}_0) dy_{1:k}\\
\nonumber& =& \int \frac{p(y_{1:k}|\mathcal{H}_1)}{p(y_{1:k}|\mathcal{H}_0)}p(y_{1:k}|\mathcal{H}_0) dy_{1:k}\\
\nonumber& =& \int p(y_{1:k}|\mathcal{H}_1) dy_{1:k} = 1 \le 1.
\end{eqnarray}
Note that we did not need to be explicit about the prior for the effect(s) under the alternative hypothesis, implying that this result holds regardless of the prior for the parameter(s) that are/is tested.

\section{Numerical check of the e-value criterion}\label{App_numericalchecks2}
\begin{figure}[htp]
\begin{center}
\includegraphics[width=13.8cm]{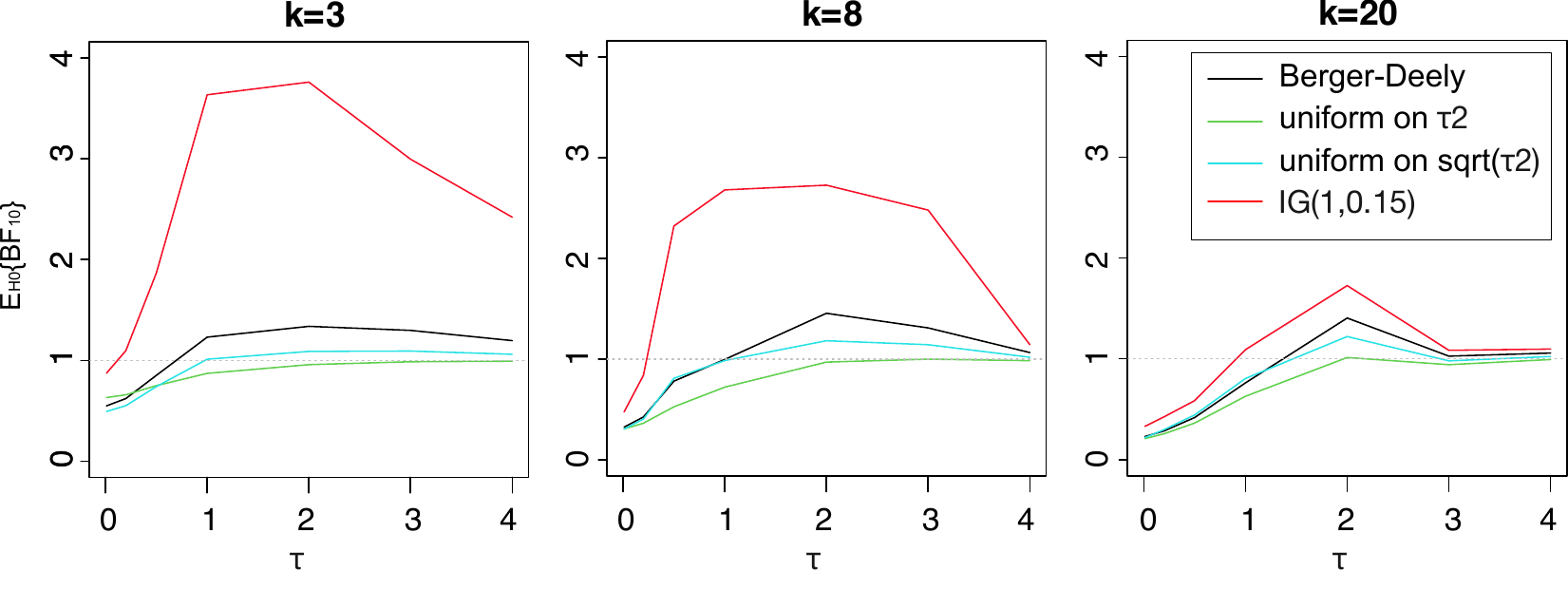}
\caption{Average Bayes factor, $B_{10}$, for two-sided test of the global effect under the random effects model based on 10,000 randomly generated data sets under $\mathcal{H}_0$ (where the global effect is zero).}
\label{fig_evalue_check}
\end{center}
\end{figure}

A simple numerical simulation was carried out to get an indication of whether the e-value condition \eqref{safetycheck} holds for the Bayes factors based on different priors of the between-study heterogeneity. Different true values of the between-study heterogeneity were considered to assess whether the condition holds over the entire range of true value for the between-study heterogeneity (which is a critical condition for an e-value). The following conditions were included: 
\begin{itemize}
\item Number of studies: $k=3$, 8 or 20.
\item Between-study standard deviation: $\tau=.01,$ .2, .5, 1, 2, 3, or 4.
\item Priors: 
\begin{itemize}
\item Improper uniform prior on $\tau^2$: $p(\tau^2)=1$.
\item Improper uniform prior on $\tau$: $p(\tau^2)=1/\sqrt{\tau^2}$.
\item Improper Berger-Deely prior \citep{berger1988bayesian}.
\item Proper inverse gamma prior: $\tau\sim IG(1,0.15)$
\end{itemize}
\item Study-specific error standard errors $\sigma_i$ were sampled from a uniform distribution: $\mathcal{U}(.2,.8)$.
\end{itemize}
Under each condition, 10,000 data sets were generated. The arithmetic average Bayes factor, $B_{10}$, was computed, which is an estimate of the expected value of the Bayes factor. Bayes factors were computed under the random effects model and the marema model. The improper uniform prior on $\tau$ under the marema model was specified as $1/\sqrt{|\tau^2|}$ to allow negative variance values. Figure \ref{fig_evalue_check} shows the results for the random effects model (the lines for the marema model were very close to the results of the random effects model and therefore omitted for clarity of the plot). Interestingly, the uniform prior on $\tau^2$ seems to satisfy the e-value criterion over the entire range. The uniform prior on $\sqrt{\tau^2}$ and the Berger-Deely prior shows slight violations. The informative inverse gamma shows most severe violations of the e-value criterion, except when $\tau$ is very small. This can be explained as this informative prior has its mode at $\tau=.14$, implying that it has good frequentist properties around the anticipated values under this prior.

\section{Analytic expression of the Bayes factor under the common effect model with normal prior}
When using a conjugate normal prior for the mean effect, $\theta\sim\mathcal{N}(0,\sigma_0^2)$, under the alternative common effect model, the posterior based on $k$ studies also follows a normal distribution with mean and variance given by
\begin{eqnarray*}
m_k =\frac{\sum_i y_i /\sigma_i^{2}}{1/\sigma_0^2+\sum_i 1/\sigma_i^{2}}\text{~ and~ }
v^2_k=\frac{1}{1/\sigma_0^{2}+\sum_i 1/\sigma_i^{2}}.
\end{eqnarray*}
and the Bayes factor would have a simple analytic form
\begin{equation}
B_{01}(y_{1:k}) = \frac{p(\theta=0|y_{1:k},\mathcal{H}_1)}{p(\theta=0|\mathcal{H}_1)}=\frac{N(0|m_k,v_k^2)}{N(0|0,\sigma_0^2)} = \frac{\sigma_0}{v_k}\exp\left\{-
\frac{m_k^2}{2v_k^2}
\right\}.
\label{BFa}
\end{equation}
In the general case of manually specified priors, the posterior would not belong to a known probability distribution, and the Bayes factor would not have an analytic expression. In that case numerical techniques are required for the computation.

Due to the simple analytical form of the Bayes factor, the evidence can easily be updated when data from a new study $k+1$ become available. One only needs to update the posterior mean and the posterior variance with the new estimate, $y_{k+1}$, and variance, $\sigma_k^2$, according to
\begin{eqnarray*}
m_{k+1} = \frac{y_{k+1}/\sigma_{k+1}^2 + m_k/v^2_k}{1/\sigma_{k+1}^2 + 1/v^2_k}\text{~ and~ }
v^2_{k+1} = \frac{1}{1/\sigma_{k+1}^2 + 1/v^2_k}.
\end{eqnarray*}
The synthesized Bayes factor \eqref{BFa} can then be computed from these updated posterior quantities.

\section{Proof that the Bayes factor is an e-value under the common effect and fixed effects model}\label{AppSafeProof}
Now we show that the Bayes factor $B_{10}$ satisfies condition \eqref{safetycheck} under the common effect model (which does not contain any nuisance parameters), implying that the Bayes factor is an e-value. The proof only requires basic probability calculus:
\begin{eqnarray}
\nonumber\mathbb{E}_{\mathcal{H}_0}\{B_{10}(y_{1:k})\} &=& \int B_{10}(y_{1:k})p(y_{1:k}|\mathcal{H}_0) dy_{1:k}\\
\nonumber& =& \int \frac{p(y_{1:k}|\mathcal{H}_1)}{p(y_{1:k}|\mathcal{H}_0)}p(y_{1:k}|\mathcal{H}_0) dy_{1:k}\\
\nonumber& =& \int p(y_{1:k}|\mathcal{H}_1) dy_{1:k} = 1 \le 1.
\end{eqnarray}
Note that we did not need to be explicit about the prior for the global mean under the alternative hypothesis, implying that this result holds regardless of the prior for the parameter that is tested.

\section{Empirical analyses using different prior distributions}\label{AppEmpEx}

In this section, we report the results on how sensitive the results of the two examples are to the specification of the prior distributions. 

\subsection{Example 1: Statistical learning of people with language impairment}

For the meta-analysis by Lammertink et al. (2017), we used as alternative prior distribution for the (average) effect a normal distribution with mean of zero and standard deviation of 0.5. This prior distribution is more informative than the previously used prior distribution (i.e., $N(0,1)$) suggesting that extreme standardized mean differences are less likely using this prior distribution. 

The results of this sensitivity analysis are reported in Table \ref{tab:results_sens_lammertink_es}. The first two columns show the results of the previously used prior distribution and the last two columns show the results of the sensitivity analysis. All methods yielded larger Bayes factors and posterior model probabilities in the sensitivity analysis implying that there was more evidence for the unconstrained model. However, the alternative prior distribution yielded the same conclusions as the previously used prior distribution, because there was strong evidence for the existence of an effect for both prior distributions.

\begin{table}[]
\caption{Results of using different priors for the (average) effect for testing $\mathcal{H}_1$ versus $\mathcal{H}_0$ for the meta-analysis by Lammertink et al. (2017). The first two columns show the results for the used default prior $N(0,1)$ and the last two columns show the results using the prior $N(0,0.5)$ as sensitivity analysis. Bayes factors (BF$_{10}$) and posterior probabilities ($PHP(\mathcal{H}_1)$) are presented.}
\label{tab:results_sens_lammertink_es}
\begin{tabular}{lcccc}
\hline
       & \multicolumn{2}{c}{Prior for $\mu$ and $\theta$: $N(0,1)$} & \multicolumn{2}{c}{Prior for $\mu$ and $\theta$: $N(0,0.5)$} \\ \hline
       & $BF_{10}$                     & $PHP(\mathcal{H}_1)$                 & $BF_{10}$                      & $PHP(\mathcal{H}_1)$                  \\ \hline
CE     & $4.41 \times 10^6$                 & 1                          & $5.7 \times 10^6$                  & 1                           \\
RE     & 138.402                       & 0.993                      & 178.242                        & 0.994                       \\
MAREMA & 251.198                       & 0.996                      & 325.649                        & 0.997                       \\
BMA    & $2.41 \times 10^3$                     & 1                          & $3.13 \times 10^3$                      & 1                           \\
FE     & $3.71 \times 10^3$                     & 1                          & $1.59 \times 10^4$                     & 1                           \\ \hline
\end{tabular}
\end{table}

Table \ref{tab:results_sens_lammertink_het} presents the results of the sensitivity analyses using different priors for the between-study heterogeneity. These analyses were only conducted for the random-effects and marema models, because the other models do not contain a between-study heterogeneity parameter or cannot deal with uninformative improper priors. In this sensitivity analysis, we included as priors for the heterogeneity the uniform prior on $\tau^2$, the uniform prior on $\sqrt{\tau^2}$, and the inverse-gamma prior with shape 1 and scale 0.15 on $\tau$. We also present the results of the Berger-Deely prior for reference and provide the results for the above used priors for the average effect. This sensitivity analysis showed that the results were robust to specification of the prior distribution for the heterogeneity. The results of both the random-effects model and marema model were hardly affected by using different prior distributions.

\begin{table}[]
\caption{Results of using different priors for the between-study heterogeneity for testing $\mathcal{H}_1$ versus $\mathcal{H}_0$ for the meta-analysis by Lammertink et al. (2017). The first four rows show the results of the different priors when the prior of the average effect is $N(0,1)$. The last four rows show the results of the different priors when the prior of the average effect is $N(0,0.5)$. Bayes factors (BF$_{10}$) and posterior probabilities ($PHP(\mathcal{H}_1)$) are presented.}
\label{tab:results_sens_lammertink_het}
\begin{tabular}{clcccc}
\hline
\multicolumn{1}{l}{}            &                            & \multicolumn{2}{c}{RE}                      & \multicolumn{2}{c}{MAREMA}                  \\ \hline
\multicolumn{1}{l}{Prior $\mu$} & Prior heterogeneity        & $BF_{10}$            & $PHP(\mathcal{H}_1)$           & $BF_{10}$            & $PHP(\mathcal{H}_1)$           \\ \hline
\multirow{4}{*}{$N(0,1)$}       & Berger-Deely               & 138.402              & 0.993                & 251.198              & 0.996                \\
                                & Uniform on $\tau^2$        & 141.090              & 0.993                & 253.217              & 0.996                \\
                                & Uniform on $\sqrt{\tau^2}$ & 139.240              & 0.993                & 249.716              & 0.996                \\
                                & $\tau \sim IG(1,0.15)$     & 143.196              & 0.993                & 250.189              & 0.996                \\
\multicolumn{1}{l}{}            &                            & \multicolumn{1}{l}{} & \multicolumn{1}{l}{} & \multicolumn{1}{l}{} & \multicolumn{1}{l}{} \\
\multirow{4}{*}{$N(0,0.5^2)$}   & Berger-Deely               & 178.242              & 0.994                & 325.649              & 0.997                \\
                                & Uniform on $\tau^2$        & 173.388              & 0.994                & 325.177              & 0.997                \\
                                & Uniform on $\sqrt(\tau^2)$ & 178.686              & 0.994                & 325.255              & 0.997                \\
                                & $\tau \sim IG(1,0.15)$     & 179.739              & 0.994                & 326.064              & 0.997                \\ \hline
\end{tabular}
\end{table}

\subsection{Example 2: Exercising after a breast cancer surgery}

For the example meta-analysis of McNeely et al. (2010), the previously used prior for the (average) effect was based on uniform priors for the success probabilities in both groups. We created an alternative prior distribution for the (average) effect by first selecting a prior distribution for both success probabilities that are peaked around 0.3. This was a beta-distribution with $\alpha = 3$ and $\beta = 5.5$. We then sampled from this distribution 100,000 probabilities for group 1 and 100,000 probabilities for group 2 and computed the log odds ratio based on these sampled probabilities. Subsequently, the parameters of a \textit{t}-distribution were estimated based on the computed log odds ratios using the same procedure as outlined in Appendix A. This yielded a \textit{t}-distribution with 41 degrees of freedom and scale of 1.067. 

Table \ref{tab:results_sens_mcneely_es} shows the results of the initially used prior distribution (i.e., $t_{13}(0,2.35)$) in the first two columns and of the alternative prior distribution in the last two columns. Using the alternative prior distribution, there is more evidence for the unconstrained model for all methods, because very extreme effect sizes are unlikely under this more informative alternative prior. However, only the Bayes factor of the fixed-effects model was slightly larger than three implying that there was at most weak evidence favoring the unconstrained model over the null model. These opposing results of the fixed-effects model using the alternative prior distribution were caused by studies three and five that have larger Bayes factors when using the alternative prior distribution compared to the previously used prior distribution (0.664 vs. 0.347 for study three and 0.548 vs. 0.287 for study five). When multiplying the Bayes factors in the fixed-effects model, these studies have less influence on the Bayes factor of the fixed-effects model in the alternative prior distribution compared to the previously used prior distribution. 

\begin{table}[]
\caption{Results of using different priors for the (average) effect for testing $\mathcal{H}_1$ versus $\mathcal{H}_0$ for the meta-analysis by McNeely et al. (2010). The first two columns show the results for the used default prior $N(0,1)$ and the last two columns show the results using the prior $N(0,0.5)$ as sensitivity analysis. Bayes factors (BF$_{10}$) and posterior probabilities ($PHP(\mathcal{H}_1)$) are presented.}
\label{tab:results_sens_mcneely_es}
\begin{tabular}{lcccc}
\hline
       & \multicolumn{2}{c}{Prior for $\mu$ and $\theta$: $t_{13}(0,2.35)$} & \multicolumn{2}{c}{Prior for $\mu$ and $\theta$: $t_{41}(0,1.067)$} \\ \hline
       & $BF_{10}$                   & $PHP(\mathcal{H}_1)$                   & $BF_{10}$                    & $PHP(\mathcal{H}_1)$                    \\ \hline
CE     & 1.112                       & 0.526                        & 2.306                        & 0.697                         \\
RE     & 0.324                       & 0.245                        & 0.641                        & 0.391                         \\
MAREMA & 0.349                       & 0.259                        & 0.698                        & 0.411                         \\
BMA    & 0.659                       & 0.397                        & 2.052                        & 0.672                         \\
FE     & 0.195                       & 0.163                        & 3.173                        & 0.760                         \\ \hline
\end{tabular}
\end{table}

Table \ref{tab:results_sens_mcneely_het} presents the results of the sensitivity analyses when using a different prior distribution for the heterogeneity. This table has the same format as shown for the first example. The results in the table illustrate that the Bayes factor and posterior probability of this example were not sensitive to using different prior distributions of the heterogeneity.

\begin{table}[]
\caption{Results of using different priors for the between-study heterogeneity for testing $\mathcal{H}_1$ versus $\mathcal{H}_0$ for the meta-analysis by McNeely et al. (2010). The first four rows show the results of the different priors when the prior of the average effect is $t_{13}(0,2.35)$. The last four rows show the results of the different priors when the prior of the average effect is $t_{41}(0,1.067)$. Bayes factors (BF$_{10}$) and posterior probabilities ($PHP(\mathcal{H}_1)$) are presented.}
\label{tab:results_sens_mcneely_het}
\begin{tabular}{clcccc}
\hline
\multicolumn{1}{l}{}               &                            & \multicolumn{2}{c}{RE} & \multicolumn{2}{c}{MAREMA} \\ \hline
\multicolumn{1}{l}{Prior $\mu$}    & Prior heterogeneity        & $BF_{10}$ & $PHP(\mathcal{H}_1)$ & $BF_{10}$   & $PHP(\mathcal{H}_1)$   \\ \hline
\multirow{4}{*}{$t_{13}(0,2.35)$}  & Berger-Deely               & 0.324     & 0.245      & 0.349       & 0.259        \\
                                   & Uniform on $\tau^2$        & 0.325     & 0.245      & 0.346       & 0.257        \\
                                   & Uniform on $\sqrt{\tau^2}$ & 0.326     & 0.246      & 0.359       & 0.264        \\
                                   & $\tau \sim IG(1,0.15)$     & 0.323     & 0.244      & 0.349       & 0.259        \\
\multicolumn{1}{l}{}               &                            &           &            &             &              \\
\multirow{4}{*}{$t_{41}(0,1.067)$} & Berger-Deely               & 0.641     & 0.391      & 0.698       & 0.411        \\
                                   & Uniform on $\tau^2$        & 0.644     & 0.392      & 0.702       & 0.412        \\
                                   & Uniform on $\sqrt(\tau^2)$ & 0.653     & 0.395      & 0.709       & 0.415        \\
                                   & $\tau \sim IG(1,0.15)$     & 0.635     & 0.388      & 0.704       & 0.413        \\ \hline
\end{tabular}
\end{table}

\end{document}